\documentclass[manuscript]{aastex}

\usepackage[utf8]{inputenc}
\usepackage{graphicx}
\let\captionbox=\undefined
\usepackage{captcont}
\usepackage[footnotesize]{caption}
\usepackage[english]{babel}
\usepackage{ amssymb }
\usepackage{ragged2e}
\justifying

\slugcomment{Not to appear in Nonlearned J., 45.}

\shorttitle{Origin of Earth's Water}
\shortauthors{de Souza Torres et al.}

\begin{document}

\title{A Compound model for the origin of Earth's water}

\author{A. Izidoro\altaffilmark{1}}

\author{K. de Souza Torres\altaffilmark{2} }

\author{O. C. Winter\altaffilmark{1}}
\email{ocwinter@pq.cnpq.br}

\and
\author{ N. Haghighipour\altaffilmark{3} }

\altaffiltext{1}{
UNESP, Univ. Estadual Paulista - Grupo de Dinâmica Orbital \& Planetologia, Guaratinguetá, CEP 12.516-410, São Paulo, Brazil}
\altaffiltext{2}{UTFPR, Universidade Tecnológica Federal do Paraná, Brazil}
\altaffiltext{3}{Institute for Astronomy and NASA Astrobiology Institute, University of Hawaii-Manoa, Honolulu, HI 96822, USA }


\begin{abstract}

One of the most important subjects of debate in the formation of the solar system is the 
origin of Earth's water. Comets have long been considered as the most likely source of the 
delivery of water to Earth. However, elemental and isotopic arguments suggest a very small 
contribution from these objects. Other sources have also been proposed, among which, local 
adsorption of water vapor onto dust grains in the primordial nebula and delivery through 
planetesimals and planetary embryos have become more prominent. However, no sole source of 
water provides a satisfactory explanation for Earth's water as a whole. In view of that, using numerical simulations, we have developed a compound model incorporating both the principal endogenous and exogenous 
theories, and investigating their implications 
for terrestrial planet formation 
and water-delivery. Comets are also considered in the final analysis, as it is likely that at least 
some of Earth's water has cometary origin.
We analyze our results comparing two different water distribution models,
and complement our study using D/H ratio, finding possible relative contributions
from each source, focusing on planets formed in the habitable zone.
We find that the compound model play an important role by showing more advantage in the amount and time
of water-delivery in Earth-like planets.

\end{abstract}


\section{Introduction}

The origin of Earth's water is one of the most outstanding questions in planetary science. 
The locations of the regions within the early solar system where water-carrying objects 
might have contributed to the accretion of Earth is a matter of intense debate.
 It is widely accepted that the protosolar nebula was hotter and denser towards its center, and cooler and less dense farther out (Encrenaz, 2006). This temperature gradient significantly affected nebula’s chemical composition and the distribution of its water and icy materials. Close to the center only metal and
silicates condensed whereas more volatile materials condensed farther out from the Sun. The first solid particles had microscopic sizes. These objects gradually stuck to
other particles and grew to larger dust grains. In an oxygen-rich region, debris
formed carbonaceous chondrites that carry up to 10 percent water (Morbidelli et al.
2000). Beyond the giant planets, water condensed in large quantities and formed
comets, which are up to 80 percent made of ice (Jessberger, Kissel \& Rahe, 1989).
.

Compared with these icy materials, Earth contains little amount of water. Only about 0.02 
percent of Earth's mass is in its oceans, and somewhat more water sits beneath its surface. 
An estimate of the amount of water inside Earth points to values ranging from 
$1 O_\oplus$ ($O_\oplus$ = mass of Earth's oceans = $1.4\times 10^{24}$ g) to 50 $O_\oplus$ 
with $\sim 10\, O_\oplus$ being the value that is generally considered as the amount of water
inside Earth's mantle (Drake \& Campins, 2006). A recent estimate of the amount of water in Earth’s interior by Marty (2012) also agrees with these values. As suggested by this author, Earth contains $\sim 4-12O\oplus$ in its interior. The important fact is that, irrespective of 
these values, Earth still has substantially more water than expected from a body at 1 AU 
from the Sun.

In the past decade many attempts were made to explain the origin of Earth's water. 
Suggested sources can be divided into endogenous and exogenous, including water adsorbed 
by dry silicate grains in the nebula (Stimpfl, Lauretta \& Drake 2004), delivery through 
asteroids, comets, planetary embryos, and planetesimals (Morbidelli et al. 2000; Raymond, 
Quinn \& Lunine 2004; O'Brien, Morbidelli \& Levison 2006; Raymond et al. 2004, 2006, 2009;
Lunine 2003; Drake \& Campins 2006), and  water production through oxidation of an hydrogen-rich atmosphere  (Ikoma \& Genda, 2006). These different sources may be distinguished by their isotopic differences such as
their deuterium-to-hydrogen (D/H) ratios as well as their predictions of the delivered amount of
water relative to the total mass. 

Among these possibilities, comets have long been considered as an attractive exogenous source 
of Earth's water. These objects that are made of ice, are believed to likely retain 
the isotopic composition that they acquired during their formation (Drouart et al. 1999). 
Whether comets were the sole source of the delivery of water to Earth is, however, uncertain. 
The measurements of the D/H ratio of water in eight Oort Cloud comets are on average twice 
that of the Standard Mean Ocean Water (SMOW) and fifteen times the value of the D/H ratio in 
the early solar nebula (Table 1). Although the original value of the D/H ratio of Earth’s water is unknown, and it is unclear how that value changed during the geophysical and geochemical evolution
of Earth (for instance, as shown by Genda \& Ikoma (2008), the Earth’s D/H ratio
could have increased by a factor of 9 had Earth had a massive hydrogen atmosphere
that experienced slow hydrodynamic scape), many researchers have used the
incompatibility between the D/H ratio in comets and in SMOW as an argument to
rule out comets as the main source of the delivery of water to Earth.

To resolve this issue, Owen \& Bar-Nun (1995) and Delsemme (1998) suggested that comets formed 
in Jupiter's region may be the source of Earth's water. These comets have lower D/H ratios 
since, compared to the comets from the Oort Cloud, they have formed in a warmer environment
[e.g., the D/H ratio of the Jupiter-family comet 103P/Hartley 2 is almost the same as 
that of the SMOW, (Hartogh et al. 2011)]. However, models of the dynamical evolution of solar system do not seem to support this idea. As shown by Morbidelli et al. (2000), assuming Jupiter and Saturn were
formed in their current orbits, cometary material delivered to Earth might have been originated from the regions beyond the orbit of Uranus as the comets formed in the Jupiter region would have very short lifetimes implying that their probability of their collision with Earth has been very low. Gomes et al. (2005) who showed that a cometary influx to Earth based on the dynamics of the outer planets delivers no more than 
$\sim 1.8\times 10^{23}$ g of material which is only about 6 percent of the current mass of 
Earth's oceans. As shown by Morbidelli et al. (2000) and Dauphas et al. (2000), comets 
constitute a minor source of Earth's water and the contribution of cometary water is 
smaller than $\sim 10\%$. It is, however, important to note that as shown by
Ipatov \& Mather (2007), if the effect of the dynamics of the outer planets 
on the scattering of Jupiter family comets are studied during the times that are earlier than 
the time considered by Gomes et al. (2005), the probability of the collision of these objects
with Earth would be large enough to justify the delivery of all Earth's ocean entirely through these
bodies during formation of the giant planets. 

Another challenge to the notion of water-delivery by comets comes from the analysis of 
noble gases and other elemental isotopic ratios. For instance, Swindle \& Kring (2001) have 
argued that based on their analysis of noble gas ratios, comets could not have been able to
supply a significant fraction of Earth's water unless either the comet delivery of water 
occurred in the first 100 Myr of Earth's history, or water was delivered by comets from 
regions other than the Oort cloud. Based on elemental and isotopic arguments, Drake \& Righter 
(2002) were able to confirm the findings of Swindle \& Kring (2001) and limited the 
contribution of comets to Earth's water to 50 percent.

Dynamical simulations and other isotopic analyses suggest that a more reliable value for the contribution of cometary water to Earth is probably smaller than 10-15 percent (Morbidelli et al. 2000; Drake \& Righter 2002). This is in agreement with the finding of Hutsem\'ekers et al. (2009) who used $^{15}N/^{14}N$ isotopic ratio and showed that the amount of cometary water delivered to Earth could not be more than 9\%. It is, however, necessary
to caution that studies based on isotopic analysis can lead to discussions as to which measured 
value of the D/H ratio in comets is a true representative of their bulk composition since the 
primary source of information about comets continues to be the studies of their comae. As shown by
Weirich, Brown \& Lauretta (2004), the D/H ratio would be expected to rise during diffusion and 
sublimation. This was also shown in an experiment by Schmidt et al. (2005) where these authors
studied comet sublimation and showed that an upward trend is observed in the D/H ratio in the 
evaporated material independent of their bulk composition. These authors argued that in order to
obtain a more realistic value of the D/H ratio in comets, it is important to understand the 
amount of bulk ice in a comet's coma.

Another possible exogenous source of Earth's water is the primitive asteroids.
The D/H ratios of the individual carbonaceous chondrites range from 1.2 $\times 10^{-4}$ 
to 3.2 $\times 10^{-4}$ (Lecuyer et al. 1998), a value that is very close to the D/H of SMOW,
implying that hydrated carbonaceous asteroids originated from the primordial asteroid belt 
might have been able to provide terrestrial planets' water. 
Water-carrying asteroids from the outer asteroid belt ($>$ 2.5 AU) 
can be scattered into the terrestrial region due to their interactions with
giant planets and planetary embryos. While this presents a viable mechanism for the delivery of water, models of the dynamical evolution and sculpting of asteroid belt indicate that its efficiency is low
and its contribution to the total water budget of Earth may be very small. The
original simulations of Morbidelli et al. (2000) in which planetesimals were
considered to be massless particles, for instance, suggest only 0.13\% of primitive
asteroids would have been accreted by Earth (Morbidelli et al. 2000). Assuming
10\% water content for a typical hydrated carbonaceous asteroid, and no loss of
mass and water in each collision, this rate of accretion requires the mass of the
asteroid belt beyond 2.5 AU to be at least 4 times the mass of Earth in order for
primitive asteroids to deliver the current amount of water to Earth (Morbidelli et al 2000).
More advanced, high-resolution simulations by  Raymond et al. (2007) show that when the planetesimals are considered to have mass, the contribution of primitive asteroids could be as high as $\sim 5\%$ of the initial population.

In these preliminary simulations  by Morbidelli et al (2000), a second issue with  primitive asteroids as the source of Earth’s water is the time of water-delivery. While the mass-requirement of water-delivery through primitive asteroids is comparable with the estimates of the original mass of the asteroid belt, the time of the delivery is not comparable with the time of terrestrial planet formation. For instance, considering primitive asteroids as massless particles, Morbidelli et al (2000) showed that the process of the delivery of the current amount of Earth’s water to the Earth’s accretion zone will take no longer than 40 Myr. At this time, Earth is young with a mass $<$ 60\% of its current value, and still forming. As Earth continues its growth, it is subject to numerous impacts by planetesimals and several major collisions with planetary embryos. It is not certain (and in fact it is even doubtful) if Earth could have retained water that it received through the impacts of primitive asteroids (Genda \& Abe 2005, Canup \& Pierazzo 2006). In the framework of this scenario, it is important to note that in more recent experiments by  Raymond et al. (2006, 2007, 2009) and O'Brien et al. (2006), where planetesimals have mass, a significant part of the delivery of asteroidal material to planets around 1 AU happens when these bodies have accreted more than 50\% of their final masses.

The requirement that the asteroid belt must have been much more massive in the past in
order to facilitate the delivery of water by asteroids, combined with the fact that 
asteroids/planetesimals and planetary embryos (moon- to Mars-sized bodies formed shortly
after the formation of planetesimals) are scattered into varieties of orbits as a result
of interacting with one another and with giant planets suggests that the delivery of water 
to Earth must have occurred during the entire course of Earth's formation through the impacts 
of hydrated planetesimals and water-carrying planetary embryos. These objects originated 
from the outer asteroid belt and deposited their water contents when impacting Earth (Morbidelli et al 2000; Raymond et al., 2004, 2006, 2007, 2009; O'Brien et al., 2006).
This model is widely accepted as the main mechanism for the delivery of water to Earth and
points to the outer asteroid belt as the reservoir of the water-carrying materials.

 The main drawback for considering asteroidal water as a source of Earth's water is that the Earth’s primitive upper mantle  has a significantly higher ${\rm ^{187}Os/^{188}Os}$ ratio than carbonaceous chondrite. In fact,  Os isotopic composition of Earth’s primitive upper mantle matches those obtained from anhydrous ordinary chondrites, and is distinctly higher than anhydrous enstatite chondrites (Drake, 2005).

It is important to note that the above-mentioned models of water-delivery are
based on the assumption that giant planets maintain their (current) orbits during the formation of terrestrial planets. Recently Walsh et al (2011) have shown that the delivery of water through hydrated asteroid and planetary embryos can also occur
during the migration of giant planets. Known as the Grand Tack model, these
authors proposed an early inward-then-outward migration of Jupiter and Saturn in
a gas-rich phase, and showed that unlike the previously mentioned models, the
water-delivery to terrestrial planet zone occurs when both planets are migrating
outward and scatter water-rich bodies initially orbiting beyond the orbit of Jupiter,
to the inner regions.

In addition to the exogenous sources explained above, it has been suggested that the 
Earth's water might have an endogenous origin: water could have come directly from the solar 
nebula. Stimpfl et al. (2004) examined the role of physisorption by modeling the adsorption 
of water on dust grains at 1000 K, 700 K, and 500 K using Monte Carlo simulations and showed 
that grains accreted to form Earth could have adsorbed 1-3 Earth oceans of water prior to their
accretion. As shown by these authors, the efficiency of water adsorption increases as the 
temperature decreases.

Although the observations of water vapor and atomic hydrogen at 1 AU around young
stellar objects such as MWC 480 (Eisner, 2007), and the presence of forsteritic olivine in
the dust clouds surrounding young solar-type stars are supportive of the water adsorption
model (Muralidharan et al. 2008; Stimpfl et al. 2006; Eisner, 2007; Bethell, \& Bergin,
2009), this scenario suffers from issues related to the retention of water during the growth
of dust grains to larger objects. Another issue with this model is the value of the D/H
ratio since the D/H ratio of the nebula is ~7 times smaller than that of SMOW. That
means, if the adsorption process had provided all the water on Earth, a mechanism would
be required to raise the D/H ratio from its solar value to that of the SMOW (Drake,
2005).

It is necessary to note that the compositional gradient of planetesimals and
planetary embryos in the solar nebula, in particular the concentration of their
volatiles depends on their thermal evolution during their growth. Several factors
contribute to the thermal evolution of a body including local heating by the Sun,
concentration of radioactive nuclides, proximity to the magnetic field of the Sun, and
the final size of the body. These processes are very complex and their study is
beyond the scope of this paper. We refer the reader to Nuth (2008) for more
details.

While, as explained above, the origin of Earth's water is still unresolved, the uncertainties
regarding the amount of water in Earth's mantle have turn this issue into a topic of debate as well.
For instance, while Hirschmann et al. (2005) argue for a highly wet mantle that
contains approximately $20 {O_\oplus}$, the model by Smyth et al. (2006) suggests
an almost dry mantle with no more than 2 Earth's ocean.  
A comprehensive model of the formation of Earth and origin of its water has to address these 
issues as well. As mentioned above, planetesimals and planetary embryos have been able to
provide a large portion of Earth's water. However, other sources such as comets, primitive
asteroids, and water adsorbed on dust grains have also had their shares and contributions. What separates
these sources from one another is the time of their operation. As indicated by geological data,
each of these events occurred at a different time during the formation of Earth. In order to be able
to address the problem of the origin of Earth's water properly, it is necessary to determine how much 
water was delivered by each of these mechanisms, and at what time during the
terrestrial planet formation. This paper presents an attempt to address these questions.
Considering that the contribution from comets is $\leq 10\%$ of the Earth's water budget, we
focus our study on water-delivery through planetesimals and planetary
embryos, and combine that with water adsorption on dust grains as a source of Earth's water. Our goal is to
answer the following question; Assuming water adsorption on dust grain as a viable mechanism
for the delivery of some of Earth's water, would the inclusion of this mechanism in the currently
accepted model of water-delivery through planetesimals and planetary embryos be consistent with 
geological constraints on the amount of water in Earth? A positive or negative result will enable
us to assess the validity of the scenario in which water adsorbed on the surface of dust grains
provides the source of Earth's water.

In the following section we discuss our model. Section 3 presents the details of our numerical
experiment, and in section 4, we analyze the results. Section 5 concludes this study
by presenting a summary and discussing the implications of its results.

\section{Model}

Since we are interested in studying the contribution of water from endogenous sources,
we consider two disk models ``A'' and ``B''.
The disk in model A is bi-modal. It consists of an inner part from 0.5 AU to 2.5 AU
where it is populated only by planetary embryos and an outer region extending from 2.5 AU 
to 4 AU where it contains only planetesimals. 
The surface density of the disk 
is assumed to have the following two-tiered radial profile

\begin{equation}
\Sigma_{\rm A}(r)=
\left\{
\begin{array}{lll}
\Sigma_1 (r/{1 {\rm AU}})^{-3/2},\hspace{.3cm} 0.5 {\rm AU} < r < 2.5 {\rm AU} \\  \\ 
\Sigma_{2} (r/{5 {\rm AU}})^{-3/2 },\hspace{.3cm} 2.5 {\rm AU} < r < 4.0 {\rm AU}.
\end{array}
\right.
\end{equation}

\vskip 10pt
\noindent
Model ``B''  is a disk of planetary embryos only, and extends from 0.5 AU to 4 AU. 
The surface density of this disk is given by 

\vskip 5pt
\begin{equation}
\Sigma_{\rm B}(r)=
\begin{array}{lll}
\Sigma_1 (r/{1 {\rm AU}})^{-3/2}, \hspace{.35cm} 0.5 {\rm AU}< r < 4.0 {\rm AU} \\
\end{array}
\end{equation}

\vskip 10pt
\noindent
In both models, planetary embryos are distributed randomly with a mutual separation of
5-10 Hill radii. At each semimajor axis, the mass assigned to an embryo is given by

\begin{equation}
M_{{\rm embryo}} = \left[\frac{2\pi \Sigma \, {a^2}}{(3M_\odot)^{1/3}}\right]^{3/2}
\end{equation}

\vskip 10pt
\noindent
where $a$ is the semimajor axis of the embryo, $\Sigma={\Sigma_{\rm A}}(a),\,{\Sigma_{\rm B}}(a)$, 
is the disk's
surface density at the position of the embryo, and $M_\odot$ represents the mass of the Sun.

The planetesimals in the outer region of the disk model A are assumed to have a mass of
0.01 Earth-masses. We distributed 200 of these planetesimals between 2.5 and 4 AU following
the surface density of the disk as given by equation (1).  All planetesimals and planetary embryos were initially in circular orbits. The orbital inclinations of these objects were chosen randomly from the range of 0.0001 to 0.001 degrees. The number of planetary embryos in disk model A is between 60 and 80 and in disk model B is between 80 - 100.

Our models also include giant planets. As shown by 
many authors (Chambers \& Wetherill 1998; Chambers 2001; Raymond et al. 2004, 2006, 2009), 
these objects play an important role in the radial mixing of the disk material,
and the final assembly and water contents of terrestrial planets. We consider two models,
one with Jupiter, and one with Jupiter and Saturn, and assume that the orbital elements of these
objects, at the beginning of our simulations, are similar to their current values.

We consider a biphasic scenario for the distribution of water in both our disk models.
We consider planetesimals interior to 0.7 AU to be dry and
assume that in the region of 0.7-2.5 AU, water is distributed according to the
(endogenous) model proposed by Stimpfl et al. (2004). In this model, the amount of water varies according
to the temperature gradient profile of the solar nebula. 
An examination of the relation between temperature and distance in the solar nebula 
as given by Clark (1998), and the relation between distance and water adsorbed 
by planetesimals and planetary embryos in the inner Solar System as given by (Stimpfl et al. 2004) 
suggests that the amount of water adsorbed by an object in the inner part of the disk is given by
\begin{equation}
W=10a-7.
\end{equation}
In this equation, $W$ is the water content adsorbed (in ocean water) by terrestrial masses and $a$ 
is the object's semimajor axis in AU. For the outer region of the disk ($a>2.5$ AU) in both models, 
we use a water content representative of carbonaceous chondrites, equivalent to 5 percent of 
the mass of each object (Raymond et al. 2004, 2005a, 2005b, 2006, 2009). Following equation (4), a body interior to 0.7 AU is dry, at 1 AU it will have 0.07\% water by mass, at 2 AU its water-mass fraction will be 0.3\%, and at 2.5 AU, 0.42\% of its mass will be water. This range of water-mass fraction indicates that the
distribution of water in our model is consistent with the amount of water carried by
S-type asteroids. These objects are the dominant bodies in the inner asteroid belt
and contain $\sim$0.1\% water by mass (Abe et al., 2000). Despite this agreement, it is still
hard to explain the presence of more anhydrous primordial bodies in the inner
solar system if the adsorption of water vapor on dust grains were an efficient
process in the solar nebula (Drake, 2005).

\section{Numerical Simulations}

We integrated our systems for different combinations of the disk surface density profile and 
giant planets configurations. In both disk models (A and B), we considered three different values for the
surface density at 1 AU ($\Sigma_1$): 6 g/cm$^2$ corresponding to the surface density of the
inner solar system in the minimum mass solar nebula (Weidenschilling 1977; Hayashi 1981;
Raymond et al. 2004), 10 g/cm$^2$ as the commonly considered
maximum surface density at 1 AU, and  8 g/cm$^2$ as an intermediate value. For $\Sigma_2$ we only 
considered 3 g/cm$^2$.

We performed 54 simulations; 36 with the disk model A and 18 with the disk model B. 
In the simulations with model A, 18 were carried out with only one giant planet (Jupiter) and 
18 included both Jupiter and Saturn. In these simulations, each disk with a different
value of $\Sigma_1$ was integrated 6 times with slightly different initial configurations for
planetesimals and planetary embryos. In the simulations with model B, 9 simulations included only
Jupiter and 9 included both giant planets. In this case, we performed three simulations 
for each value of $\Sigma_1$ each with slightly different initial configurations.

Simulations were carried out for 300 Myrs using the hybrid integrator in the N-body 
integration package Mercury (Chambers 1999). The time step of simulations was set to 5 days.
We assumed all collisions to be perfectly inelastic and conserve linear momentum.

Figures 1 and 2 show the final results for all 54 simulations. The labels on the vertical
axes represent the number of the simulation, the giant planet(s), and the values of  $\Sigma_1$.
The color of each object corresponds to the value of its water-mass fraction based 
on the color coding scheme at the bottom of each figure. 
The orbital eccentricity of each object is shown by a horizontal line corresponding to the range 
of the variations in the heliocentric distance of the body from its perihelion to aphelion. 
Table A1 in the appendix shows the corresponding values of the
mass, semimajor axis, orbital eccentricity, and water content of each body. As a point of
comparison, the inner planets of the solar system and their water contents are also shown.
The gray area shows the habitable zone of the Sun from the model by Kasting et al. (1993).

An inspection of the results show a great diversity in the mass, water content,
and orbital configurations of the final objects. Figure 3 shows the initial and final values of
the total masses of the planetesimals and planetary embryos in each disk model. As shown
here, in all simulations, the disk lost a great portion of its mass. This is an expected result
that is due to a combination of ejection (i.e., reaching a heliocentric distance of
100 AU or larger), and collision with Sun or the giant planets. Tables 2 and 3 show the fraction
of the mass lost by each mechanism. As shown here, most of the mass of the disk was lost as a result
of the ejection of objects out of the system. The ejected mass is larger in systems with two giant
planets. Almost always more than 50\% of the initial mass of the disk was ejected in these systems.
This is due to the fact that in such systems, in addition to the direct gravitational perturbation
of giant planets, secular resonances such as $\nu_6$ excite the orbits of planetesimals and
planetary embryos and cause many of them to be scattered out of the system. We refer the reader
to Levison \& Agnor (2003), Raymond et al. (2009) and Haghighipour et al. (2013) for an analysis of 
the effect of $\nu_6$ on the scattering of
materials from a protoplanetary disk and the onset of terrestrial planet formation. At the end of our simulations, the mean change in the semimajor axes of Jupiter and
Saturn were $\delta_{a_J}\sim 0.03-0.05$ and $\delta_{a_S}\sim 0.01-0.03$, respectively. These values are
consistent with those reported by Raymond et al. (2004).

The strong perturbations of Jupiter and Saturn combined with the effect of secular resonances
in ejecting material from the disk in simulations where both these two planets were included 
results in smaller number of collisions in these systems compared to those simulations that
included only Jupiter. For instance, in the simulations with disk model A, the total 
massloss due to collision with  Jupiter is $\sim 0.2 {M_\oplus}$ whereas in the simulations
with the same disk when both Jupiter and Saturn are included, this value drops to $0.06 {M_\oplus}$.
Same results have been obtained for simulations with disk model B. The situation is, however,
different for the collisions with the Sun. In this case, the amount of the mass lost 
due to the collision with the Sun is higher in systems that include Jupiter and Saturn. 

An interesting result of our simulations is the connection between the values of the 
mass, semimajor axis and orbital eccentricity of the final objects, and the disk surface
density profile and giant planet(s) configurations. Figures 4 and 5 show the results in a 
mass-semimajor axis diagram. In these figures, each box represents the final results of all 6 simulations 
that correspond to its mentioned value of the surface density at 1 AU and giant planet's orbit.
As shown here, larger planets
are formed in the region between 0.5 AU and 1.5 AU and their mass distribution peaks at approximately 
0.7-0.8 AU. This is an expected results that attributes to the fact that in the inner regions, 
the perturbing effect of giant planets is negligible and the dynamics of planetesimals and planetary 
embryos is primarily driven by the interactions of these bodies with one another. A consequence of 
the latter is that in a disk with a larger surface density (where more material is available in 
its inner region), the collision and accretion of planetesimals and planetary embryos result
in a few but larger objects (Raymond et al. 2004, 2005b). In the outer region $(>1.5$ AU), 
as shown by these figures, the perturbation of giant planets causes the disk material to 
be scattered out of the system and as a result, the final objects formed in these regions carry 
less mass and are smaller. This can, for instance, be seen in the simulations of model B with 
$\Sigma=6g/cm^2$  where because of the effects of Jupiter and Saturn, only a few objects smaller 
than Earth survived the integrations.

\subsection{Water content}

Figures 4 and 5 also show the water-mass fraction of the final objects. It is important
to note that the final water content of a body, in addition to 
the contributions from water-carrying planetesimals and planetary embryos initially orbiting past 2.5 AU (hereafter refer to 
as {\it asteroidal water}), depends also on the amount of the water that was initially adsorbed in
the inner regions of disk as Eq. 4 (simulations of model B included
only planetary embryos). Results of the 
simulations of the two sets of model A (Tables 4 and 5) 
indicate that the percentage of the asteroidal water received by planets
formed around 1 AU varies between 0\% and 74\%. Despite the large amount of water
that was initially available in the disk, some of the final objects in this model
received different amounts of
water even as small as $1 O_\oplus$, the minimum value of Earth's water content.
This is due to the fact that many water-carrying bodies, especially those in the outer region of the disk,
were dynamically excited by giant planets and were scattered out of the system. 

An interesting result obtained from the simulations of model A is that despite the different
amounts of asteroidal water and mass that were received by final planets, the percentage of the asteroidal 
water in these objects in simulations with one or two giant planets is the same. 
As shown by Table A1, the amount of asteroidal
mass received by most of these planets is smaller than 10\%. Figure 4 shows that similar
to the mass-distribution of the final objects, the planets with more water are within 0.5-1.5 AU 
with their water-mass ratios peaking at 0.7-0.8 AU.

Table A1 also
shows small planets with almost no asteroidal water implying that these objects accreted only a few or no planetesimals that were orbiting initially past 2.5 AU. Naturally, these objects would be expected to have formed in the inner 
region of the disk where the amount of asteroidal water is small or zero. However, our simulations 
show that objets without asteroidal water could also form at semimajor axes larger than  2.5 AU. 
Probably these bodies were scattered to the outer region of the disk
by interacting with embryos from the inner part remaining with no contribution of water from asteroidal bodies. See Haghighipour \& Scott (2012) for the possibility of scattering of planetesimals from
inner orbits to outer regions.

Note that the cases 24A-JS and 13A-J (Table A1) have indeed some amount of water.
However, the amount of water in these bodies is smaller than 0.05 Earth's oceans. Because of
such a small amount of water, and because in the table, we only considered one digit
after the decimal point in our entries, the water contents of these bodies do not
show. The planet in the system 3B-J (Table B1), on the other hand, is actually dry.
That is because this object was initially placed between 0.5 AU and 0.7 AU which is
a dry region in our compound model.

\subsection{Water in the Habitable Zone}

We followed Kasting et al. (1993) and considered a region for the habitable zone (HZ) of the Sun 
extending from 0.9 AU to 1.37 AU. In all our 54 simulations, a total of 39 planets formed inside the HZ. 
Tables 4, 5, and 6 show the final results. One can see from these tables that the masses of
these planets vary between $\sim0.2M_\oplus$ and $\sim2.2M_\oplus$ with a water content 
ranging from $1O_\oplus$ to $38 O_\oplus$. The maximum water content of the planets formed
in the HZ did not exceed the maximum value of $50 O_\oplus$ expected for Earth (Abe et al. 2000).
In the following, we present a detailed discussion on the amount of water carried by the final
planets, the time of the delivery of that amount of water, and use the D/H to determine the amount
of cometary material that needs to be delivered to the final planets in order for the D/H of their
water contents to match that of Earth. To portray the effect of the initial water adsorption, we also make a comparison between the results of our model and	those adopting the model used by Raymond et al. (2004; 2006; 2007; 2009).

\subsubsection{Water content of the final planets}

When including only Jupiter, most planets in the simulations of models A and B were able to carry at least 
5 Earth oceans of water (Tables 4-6). However, as shown in Table 7, the efficiency of the delivery 
of more than 10 Earth oceans, which by some estimates is the total amount of water carried by Earth (Drake et al. 2005), dropped significantly in our simulations when we considered the water distribution model of Raymond et al. (2004, 2006, 2007; 2009). In the latter case, only 2 of the 13 planets ($<20\%$) that were formed in the HZ received at least 10 Earth's oceans. As a point of comparison, our compound model (Table 4) 
which included water adsorption as well, was able to deliver 10 Earth's oceans in 11 of the 13 planets formed ($>80\%$) in HZ.

A comparison between the results in Tables 4 and 5 shows that in the simulations of set
A, the inclusion of Saturn significantly reduced the total amount of water in the final
planets that formed in the habitable zone. However, this reduce in efficiency was still less
than that in models in which the water was delivered only through collisions of planetesimals and
protoplanetary bodies (Raymond's model), and did not include adsorbed water. As shown in Table 7, in
our compound model, even when Saturn is included, 12 out of 14 planets in the HZ
received more than 5 Earth’s oceans. When considering the model without water
adsorption (in our same simulations), this number dropps to only 2 out of the same 14 planets. This reduction in efficiency in both models is due to effect of secular resonance with Saturn. This resonance
increases the eccentricity of small water carrying bodies in the region around 2.2 AU, and
causes many of these objects to be ejected from the system or collide with the Sun in a
short time.

Interesting results were obtained in simulations of model B. In these simulations, 
despite that our model included water adsorption, no difference was
observed in the amount of the water carried by the final planets when
Saturn was included in the simulations. The delivery of at least 5
Earth's oceans was observed in roughly 50\% of planets that formed
in the HZ. This could be attributed to the fact that in these simulations,
the disk model B contains only planetary embryos which, unlike planetesimals,
are not easily excited by Saturn's $\nu_6$ secular resonance [the
embryo-embryo interactions around 2.0-2.5 AU in these simulations
neutralize the effect of the  $\nu_6$ resonance. See Levison \& Agnor (2003)
and Haghighipour et al. (2012) for more details]. 
Since the embryos past 2.5 AU carry 5\% water, a single collision with a proto-planet in
HZ can deliver substantial amount of water to a planet in that region.

\subsubsection{Time of water delivery}

The time of water-delivery also showed dependence on the giant planets configuration. For instance, in the simulations of the disk model A considering our compound model (model A-J, see Table 7), it took on average 19 and 36 Myr to deliver 2 and 5 Earth's oceans to the final planets in the HZ (Tables 4 and 7), respectively. This is while in our same simulations using a model of water distribution as in Raymond et al (2004, 2006, 2007; 2009), where water adsorption is not included, this time is approximately 26-117 Myr.
When Saturn was included (model A-JS, see Table 7), as expected the time of delivery was prolonged. In our simulations (Tables 5 and 7), the timescale required for a planet in HZ to receive 2-5 Earth's oceans was on average 40-56 Myr using our compound model. In models without water adsorption, this time extends to 52-181 Myr. These results indicate that in general, our model is more efficient in delivering water
during the entire course of the planets' accretion. For instance, planets formed
inside the HZ from set A suffered their last giant collision on average after 150 Myr
with a range varying from 30 Myr to 290 Myr (Table 7).

While in the model without water adsorption, only 2 planets received  $10 O_{\oplus}$
on average within 90 Myr, and no planet received more than $15O_{\oplus}$, in our
compound model, 13 planets received $10O_{\oplus}$ in about 100 Myr and 4
planets received $15O_{\oplus}$ on average in 146 Myr. Evidently, our compound
model is more efficient in delivering water to many planets during the formation of
these objects. We recall that the timescale for the formation of Earth is between 50 Myr and 150 Myr
(Touboul et al. 2007).

 Similar results have been obtained in our simulations of disk model
B (Tables 6 and 7). For instance, simulation of 12B-JS in Table 6 produced a planet with
2.26 Earth’s masses carrying a total 38 Earth’s oceans. From this amount of water, 17
Earth’s oceans were strictly due to local adsorption of water in the primordial solar
nebula. As shown in the table, water-delivery through planetesimals and planetary
embryos stopped after 31.2 Myr.

In regard to the time of the delivery of water to Earth, it is important to determine what fraction of the delivered water could have 
been retained by the Earth if the water-delivery occurred very early on when the Earth was only half of its current mass 
(Morbidelli et al. 2000). 
To answer this question, we first compare the time of water-delivery for the delivery of first two oceans ($<2 O_\oplus$)
for both models of water distribution (i.e., with and without water adsorption). For the simulations of the disk model A where only 
Jupiter was included, the delivery of the first 2 oceans occurred for both models of water distribution between 10  and 35 Myr 
in 50\% of planets formed in the HZ. As shown in Table 7, this timescale is between 0.5 and 62.1 Myr for
our water distribution model and between 2.7 and 62.1 Myr for models in which water adsorption is not included.
The close similarity between these timescales indicates that the time of the delivery of the first two oceans in both models
is statistically indistinguishable. In fact, in several of our simulations, the timescale for the delivery of the first 2 Earth's oceans 
to planets in the HZ was equal in both water distribution models. This result implies that our compound model does not play
a significant role in the early delivery of water. Given the statistical indistinguishability of the results of both models,
we conclude that both water distribution models can result in similar loss of water in the early  accretion of planets in the HZ. However, as the accretion of these objects proceeds, the compound model gains an advantage in the amount and time of water-delivery compared to models of water distribution as that in  Raymond et al (2004, 2006, 2007, 2009) and O'brien et al (2006).

 Our compound model is also capable of delivering much more water to many more
planets in a timescale consistent with the mean-time of the last giant collision. Table
7 shows the results. As shown in this table, for several planets in both models, water
accretion reaches its maximum after these planets grew to 60\% of their final
masses. The mean-time of 60\% accretion and the range of the time of the 60\%
accretion are shown in the table as well. In some cases, more than 50\% of the final
water content of planets was delivered during this final phase of planet growth.
Similar results have also been reported by O’Brien et al (2006), Walsh et al (2011),
and Morbidelli et al (2012). This result is particularly important in connection with
geochemical models of the early differentiation of Earth. As proposed by Wood et
al. (2008) and Rubie et al. (2011), the oxidation state of Earth continued
progressively as the Earth was growing. The most successful models of the Earth's core formation
require that the final 30-40\% of Earth’s accreted mass to be much more oxidized
than the material accreted earlier (Rubie et al., 2011). This suggests that the water
was added together with moderate level of volatile elements during the final 30-40\%
of Earth’s accretion and it does not take into account the amount of water that
could have been adsorbed on dust grains in the early solar nebula (Rubie et al,
2011). Our results (Table 7) also have direct implication to the recent finding by Marty (2012) who suggested that the accretion of volatile elements occurred during the main Earth-forming process instead of being a contribution from a late veneer event.

\subsubsection{Implications of D/H ratio}

As mentioned in the introduction, the primary reason for comets not being considered as the main source
of Earth's water is the contrast between the measured value of the D/H ratio of water in some
of the comets with that of SMOW. When comparing these two values, it is assumed that the ratio
of deuterium to hydrogen in comets is identical to its primordial value and the D/H ratio of SMOW 
is also representative of the time when water was delivered to Earth by comets through a late
veneer process. Although these two assumptions are not entirely valid (radiations and space weathering 
could have affected the chemical compositions of comets during their approach to Earth, the chemistry 
of SMOW could have changed due to the biological processes resulted from the evolution of life, 
and the collisions of comets with Earth could have also affected the chemistry of the material 
that was transferred to Earth from the impactor), our treatment of collisions between two objects
(i.e., perfectly inelastic with the complete transfer of all material from the impactor to the impactee
with no loss of water) allows us to evaluate the amount of cometary water
that would be necessary to raise the D/H ratio of the final objects in the HZ to that of SMOW. In doing so, we consider the value of the D/H ratio of a planet, $({\rm D/H})_{\rm Planet}$, to be given by a combination of the D/H ratio of asteroidal water, $({\rm D/H})_{\rm Ast}$, and the D/H ratio of the water adsorbed from the primordial nebula, $(\rm {D/H})_{\rm Neb}$. The cometary water contribution is not
considered at this stage and is only included afterwards in order to estimate the amount that is needed to increase the D/H ratio of water in the planet to the current value of SMOW. The following equation presents a simple relation between the quantities $({\rm D/H})_{\rm Ast}$, $(\rm {D/H})_{\rm Neb}$ and $({\rm D/H})_{\rm Planet}$
\begin{equation}
K {(\rm {D/H})_{\rm Ast}} + (1 - K) {(\rm {D/H})_{\rm Neb}} = {({\rm D/H})_{\rm Planet}}.
\end{equation}
\noindent
The value of the D/H ratio of asteroids, $({\rm D/H})_{\rm Ast}$,
varies between $1.2\times 10^{-4}$ to $3.2\times 10^{-4}$ (Lecuyer et al. 1998). Adopting the value
used in our simulations,  $({\rm D/H})_{\rm Ast}= 2.2\times 10^{-4}$, and using the D/H 
ratios of SMOW and nebula as given in Table 1, Eq. (5) suggests that in order
for the asteroidal water to complement the value of the D/H ratio of the nebula
and raise the D/H ratio of water in the final planets to the current value of 
SMOW, 64\% of the amount of the water in these objects has to be delivered  by
asteroids. In other words, a planet that receives 64\% of its water from asteroids (and,
therefore, the remaining 36\% from water adsorbed in the solar nebula) will have a
D/H ratio similar to that of Earth (SMOW). Such a planet will \textit{not} need any
cometary water to raise the value of its D/H ratio. Our simulations show that only three planets in model A (simulations 12A-J, 13A-J, 19A-JS, see Tables 4, 5 and A1) received more than 64\% asteroidal water and their
D/H ratio became higher than that of SMOW. Other interesting cases are, for instance,
planets produced in simulations 4A-J, 23A-JS, 27A-JS whose water contents, semimajor axes
and orbital eccentricities are close to those value of Earth (Tables 4, 5 and A1), and
they still need approximately 2.84\%, 11.03\% and 20.45\% cometary water in order for 
their D/H ratios to reach to that of SMOW.

The simulations of model B also present interesting results. From the total of 39 planets
that formed in the HZ, 11 were from these simulations. Table 5 shows the results. Since in the
simulations of model B, the disk contained only planetary embryos,
the cometary water necessary to raise the D/H ratio of these planets to that of SMOW varied between 
0\% and 45\%. As shown by Table 5, in four simulations, the final planets had water contents with
D/H values larger than that of SMOW (simulations 4B-J, 9B-J, 15B-JS, 17B-JS, see Tables 6 and B1).

An interesting result, when focusing only on the final planets around 1 AU (model A) is that the
D/H ratio is higher in simulations with only Jupiter compared to those where both Jupiter
and Saturn were included. This can be related to the strong effect of Saturn, in particular the
secular resonances, in dynamically exciting the orbits of planetesimals and planetary embryos which 
affects the efficiency of radial mixing from the outer region of the disk to the inner parts. Results of our simulations also indicate that despite a large amount of cometary material that is needed in a few cases to increase the D/H ratio of the final planets in the HZ to that of SMOW (Tables 4, 5 and 6), the amount of necessary cometary water in some cases agrees very well with previous studies [e.g., $\leq 10\%$ 
as in Morbidelli et al. (2000), up to 12\% as in Deloule, Robert \& Doukhan (1998), and up to 15\% 
as in Owen \& Bar-Nun, (2000)] giving relevance to locally adsorbed water. 

Although the calculated amount of the cometary water necessary to increase the
D/H ratio of the final planets in the simulations of models A and B agree with 
estimates from other works, an important questions would be whether models of the dynamical evolution of 
the solar system would be able to accommodate the delivery of this amount of cometary material to Earth. 
As can be seen from the results shown in Tables 4 and 5, adding $0.06 O_\oplus$ of cometary water 
as estimated by  Gomes et al. (2005) 
or even  $\sim 5\times 10^{-5}M_\oplus$ of cometary material as calculated by Morbidelli et al. (2000) 
to the final planets of these simulations in the HZ would not change the value of their D/H ratios
significantly. If we also consider the loss of water due to the impact of bodies, which according to
Marty \& Yokochi (2006) could be as high as 20\%, the situation becomes even worse. An interesting result, however, is that as shown by the simulations of model B using our compound model, it is still possible
to form planets almost only carrying asteroidal water ($\sim$ 80\%) and with D/H ratios similar to that of SMOW. Simulations show that the majority of such objects will obtain their water from planetary embryos that originally resided
at distances beyond 2.5 AU. The latter implies that the initial distribution of water in the protoplanetary 
disk plays an important role in the final D/H ratios of terrestrial planets in the HZ.

\section{Concluding Remarks}

Assuming that Earth’s water had likely more than one source, we studied the
contribution of the water locally adsorbed from the nebula to the water contents
and D/H ratio of terrestrial planets around the Sun. We carried out 54 numerical
simulations considering locally adsorbed and asteroidal water, and analyzed the
final distribution of mass and water content of the planets formed in the Sun's HZ.
Results of our simulations indicated that large planets with large amounts of
water can form in the region between 0.5 and 1.5 AU. Results also suggested that
our compound model seems to be more efficient in the amount and time of the
delivery of water over a model in which water is transferred through mere
collisions of protoplanetary bodies.

We also studied the D/H ratios of the final planets formed in the Sun's HZ and determined
the amount of the cometary material that would be required to raise the values of their
D/H ratios to that of SMOW. Results indicated that using our compound model of water distribution and assuming the D/H ratios of comets are representative of their primordial values, on average 20\% (individually, in some cases, less than 10\%) of cometary water would be necessary to raise the values of the D/H ratio in the final planets to that of SMOW. However, models of the dynamical evolution of the solar system do not seem to be able to deliver this amount of cometary material to the HZ of the Sun. For instance, as shown by Morbidelli et al. (2000), comets can only provide up to 10\% of the Earth’s
water. Our simulations indicate that such a lack of cometary material presents no barrier
to the formation of terrestrial planets with orbital elements, water contents, and D/H
ratios similar to those of Earth. Several of our simulations were able to produce Earth-like
planets in the HZ.

In carrying out our simulations, we made certain simplifying assumptions. In order to avoid complications 
with breakage and fragmentation of colliding bodies, we assumed that all collisions were perfectly inelastic. 
We also assumed that in a collision, the mass and the amount of the water of the final object would be equal 
to the sum of the masses and 
water quantities of the impacting bodies. As pointed out by Haghighipour \& Raymond (2007), this is an assumption that 
sets an upper limit for the water budget of terrestrial planets and ignores the loss of water due to the impacts 
(Marty \& Yokochi 2006) or hydrodynamic escape (Matsui \& Abe 1986). In a more realistic model, the loss of volatile 
materials in large impacts has to be taken into consideration (Genda \& Abe 2005, Canup \& Pierazzo 2006). 

Our assumption on the temperature profile of the disk is also one of the limitations of our study.
The amount of the adsorbed water onto the solar nebula grains and the distribution of asteroidal water in the protoplanetary disk
strongly depends on the disk temperature profile. In our simulations, we followed the model presented by Clark (1998).
However, there is a rich literature on disk temperature profile and models of water adsorption in accretion disks 
(e.g., Boss 1996, Clark 1998, Sasselov \& Lecar, 2000, Fegley 2000, Muralidharan et al. 2008, Albarede, 2009). 
The choice of different temperature profile will clearly result in planetary bodies with different 
amounts of water. The analysis of the connection between the disk temperature and the water contents of the final
terrestrial planets is outside the scope of this paper, and will be presented in future articles.

Another limitation of our study is its low resolution. Since the speed of an $N$-body integration is proportional to $N^2$,
in order to keep the time of our simulations in a manageable level, we limited our integrations to only a few hundred objects.
As such, the results may not be able to reveal detailed characteristics of the terrestrial planets of our solar system. For instance, 
our simulations are not able to reproduce the small eccentricities of these objects. As shown by Agnor, Canup \& Levison (1999) 
and Chambers (2001), low-resolution integrations can produce the main general properties of the final assembly of the planetary 
bodies, however, high resolution simulations such as those by O'Brien et al. (2006) and Raymond et al. (2006, 2007, 2009) 
are necessary in order to reproduce detailed dynamical properties of planets, such as the small eccentricities of Earth and Venus.

In general, our simulations produced between two and four planets (a $<$ 2AU) on well-
separated and stable orbits. Venus-size planets are formed in most of our simulations orbiting at around 0.5 AU. Several of our simulations have produced a planet inside the habitable zone of the system in a timescale of 50-150 Myr, consistent with the expected time of the formation of Earth (Jacobsen 2005; Touboul et al 2007). Mercury-size planets did not form in our simulations due to our choices for the initial individual masses of planetary embryos distributed in the inner part of disk and the position of the inner edge of protoplanetary disk (Hansen, 2009). Such difficulties (e.g., the formation of Mars- and Mercury-size objects and the
architecture of the asteroid belt) have been known to exist in the simulations of terrestrial planet formation using similar initial conditions (e.g., Wetherill, 1991; Morbidelli et al., 2000; Chambers, 2001; Raymond et al., 2004; 2006; 2007; 2009; O’Brien et al., 2006). Only recently efforts have been made to develop models that address these
difficulties (Hansen 2009; Walsh et al 2011, Izidoro et al. 2013). High-resolution simulations would be necessary to test our compound model against these complexities as well.

As discussed in this paper, no sole source of water provides a satisfactory
explanation for the origin of Earth’s water. The main argument against the water
adsorption process (and our compound model) as a mechanism for contributing to
the Earth’s water lies in the association of different classes of meteorites to different
taxonomic type of asteroids, and the connection between their corresponding water
contents and their heliocentric distance. Carbonaceous chondrites are associated
with C-type asteroids, which are primarily beyond 2.8 AU. Ordinary chondrites
with 0.1\% of their mass as water are considered fragments of S-type asteroids,
which are between 2 and 2.5 AU. Enstatite chondrites which are very dry with only
0.01\% water (Abe et al, 2000) are linked to E-type asteroids at 1.8 AU. All this
indicates that the building blocks of Earth around 1 AU must have been extremely
dry. However, as proposed by Drake (2005), it is uncertain whether the parent
bodies of ordinary meteorites were indeed anhydrous, or if these meteorites were
derived from the metamorphosed outer parts of hydrous asteroids. A deep
understanding of the relationship between meteorites, their parent bodies, and
taxonomic type of asteroids is extremely important to improve models of origin of
Earth’s water. We believe that missions such as Dawn and OSIRIS-REx will be able
to clarify this point.

Finally, our assumptions on the D/H ratio of comets and that of SMOW present another limitation of our study.
We assumed that the current D/H ratio of Earth's ocean water is representative of its primordial value. However,
this quantity could have changed over the time due to biological (life and its evolution) and chemical processes. 
Campins, Swindle \& Kring (2004) pointed out that the processes involved in planetary accretion such as degassing, 
and the evolution of hydrosphere and atmosphere are complex and may have fractionated the chemical and isotopic signature 
of the source(s) of water. There are also debates on whether the estimated bulk D/H ratio and bulk abundance of water on 
Earth is indeed truly known (Abe et al. 2000; Drake and Righter 2002; Smith et al. 2006; Genda \& Ikoma 2008). We assumed
that the D/H ratio of SMOW was primordial because of the limited knowledge of the evolution of D/H ratio in Earth's water, 
its relation to its primordial value (Williams \& Hemley, 2001), and the lack of a sizable statistical sample of the D/H ratio 
of water in other bodies of our solar system. To improve the analysis presented here requires wide understanding of the 
D/H ratio in comets, 
asteroids and meteorites, a detailed model of the variation of D/H ratio during collisions between these bodies, as well as the modeling of 
atmospheric escape and fractionation in Earth's core (Villanueva et al. 2009). Measurements of the D/H ratio in a larger sample  
of comets with HIFI (Heterodyne Instrument for the Far Infrared) on Herschel space telescope 
can also help constraining Solar nebula models (Hartogh et al. (2009).

\section{ACKNOWLEDGEMENTS}
We thank an anonymous referee, for his constructive comments that greatly
improved our manuscript.
 This work was funded by CAPES (Coordenação de Aperfeiçoamento de Pessoal de Nível Superior), CNPq
 (Conselho Nacional de Desenvolvimento Científico e Tecnológico) and FAPESP (Fundação de Amparo a
 Pesquisa do Estado de São Paulo). NH acknowledges support from the NASA Astrobiology Institute under 
Cooperative Agreement NNA09DA77A at the Institute for Astronomy, University
of Hawaii, and NASA EXOB grant NNX09AN05G.

\clearpage
\begin{figure}
\begin{minipage}{170mm} 
{\centering
\includegraphics[width=1\linewidth]{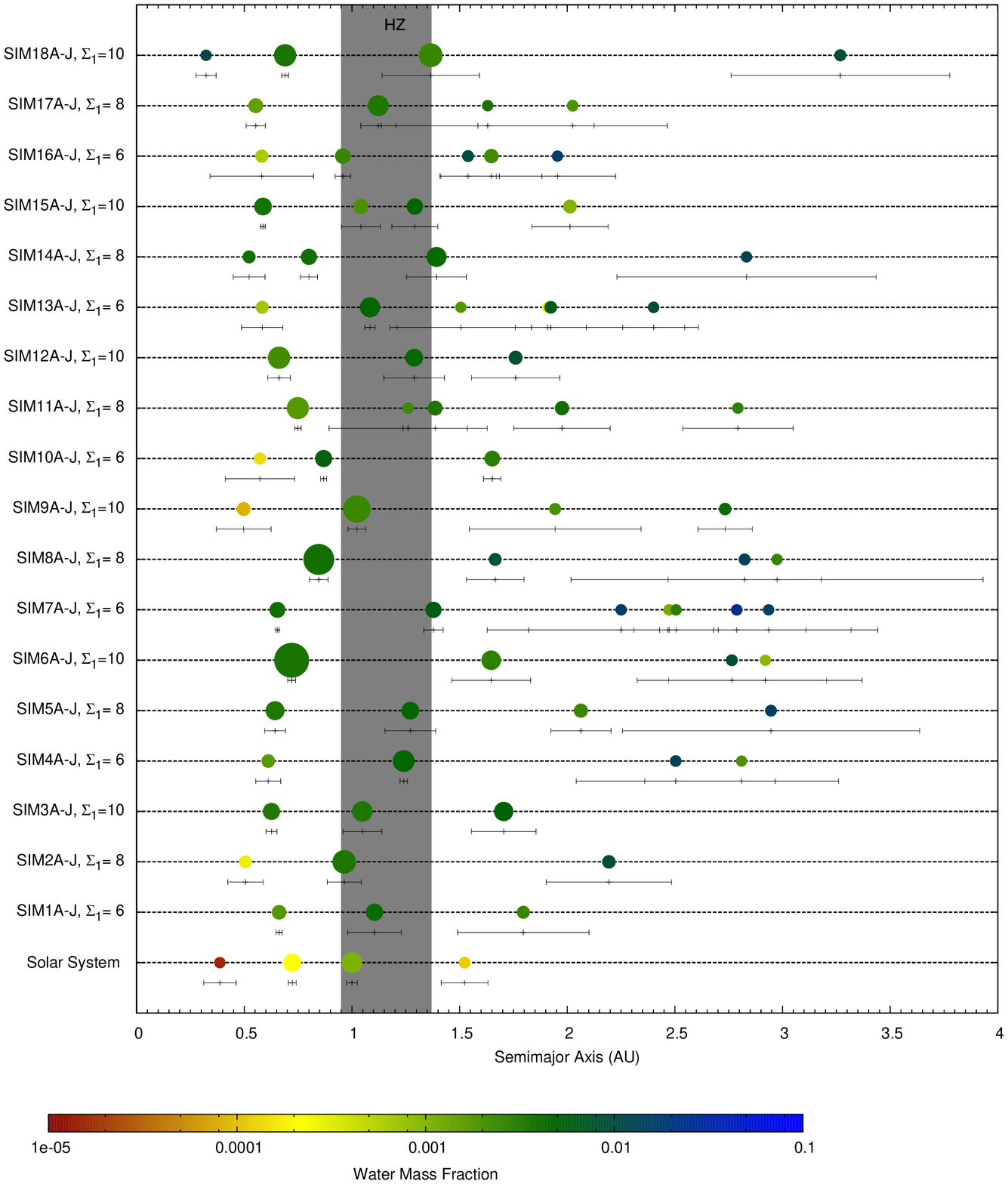}}
\label{fig1a}
\captcont{Final orbital configurations of 18 simulations of model A considering only Jupiter. The 
values of the disk surface density at 1 AU are shown on the vertical axis. The solar system is also shown for 
a comparison. The size of each body corresponds 
to its relative physical size, however, it is not to scale on the $x$-axis. The color of each planet represents 
its water-mass fraction. The eccentricity of each body is represented by a horizontal line depicting
its variation in its heliocentric distance.}
\end{minipage}
\end{figure}

\clearpage
\begin{figure}
 \begin{minipage}{170mm} 
\centering
\includegraphics[width=1\linewidth]{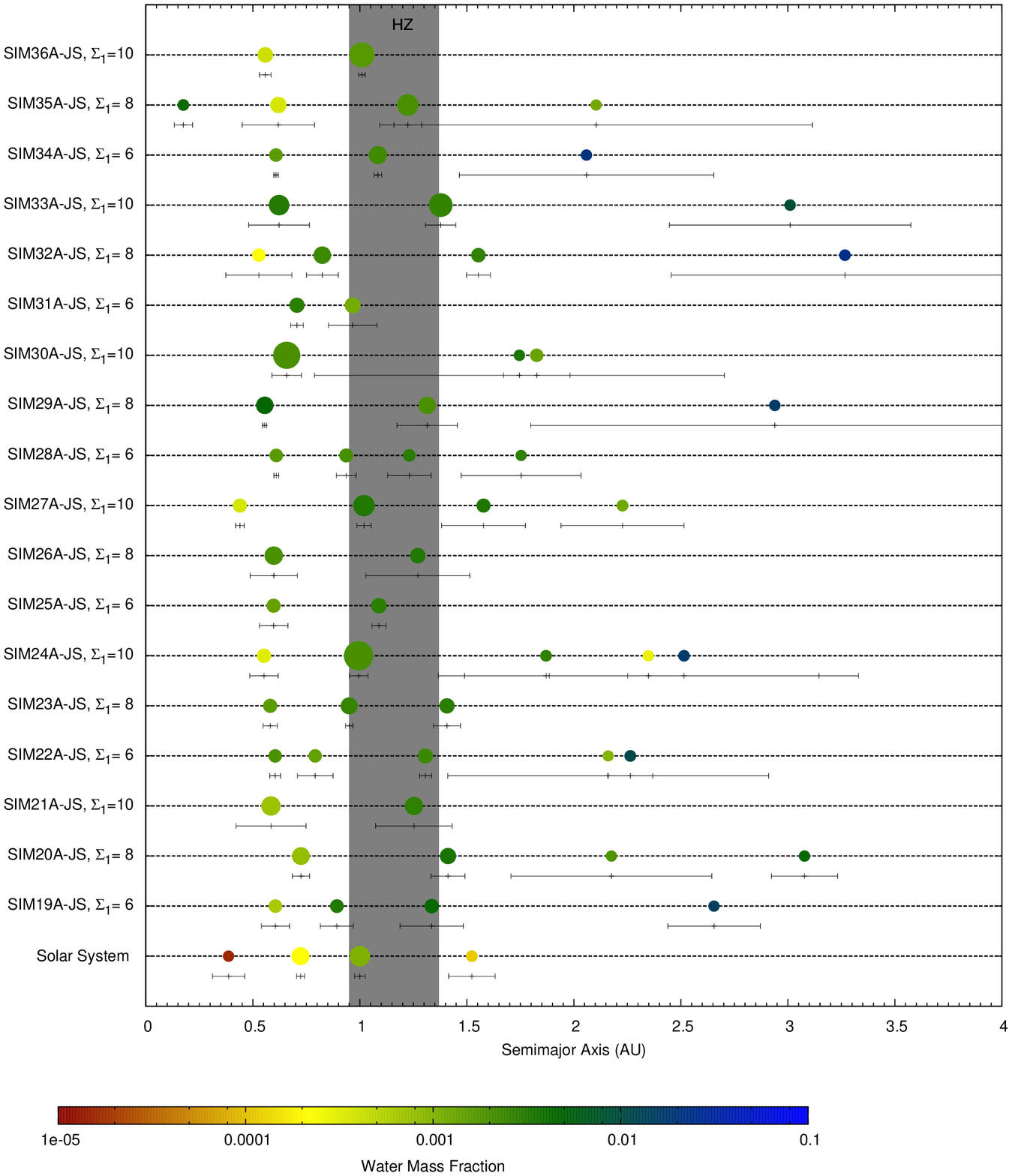}
\label{fig1b}
\caption{ Final orbital configurations of 18 simulations of model A considering Jupiter and Saturn. The 
values of the disk surface density at 1 AU are shown on the vertical axis. The solar system is also shown for 
a comparison. The size of each body corresponds 
to its relative physical size, however, it is not to scale on the $x$-axis. The color of each planet represents 
its water-mass fraction. The eccentricity of each body is represented by a horizontal line depicting
its variation in its heliocentric distance.}
\end{minipage}
\end{figure}

\clearpage
\begin{figure}
 \begin{minipage}{170mm} 
\centering
\includegraphics[width=1\linewidth]{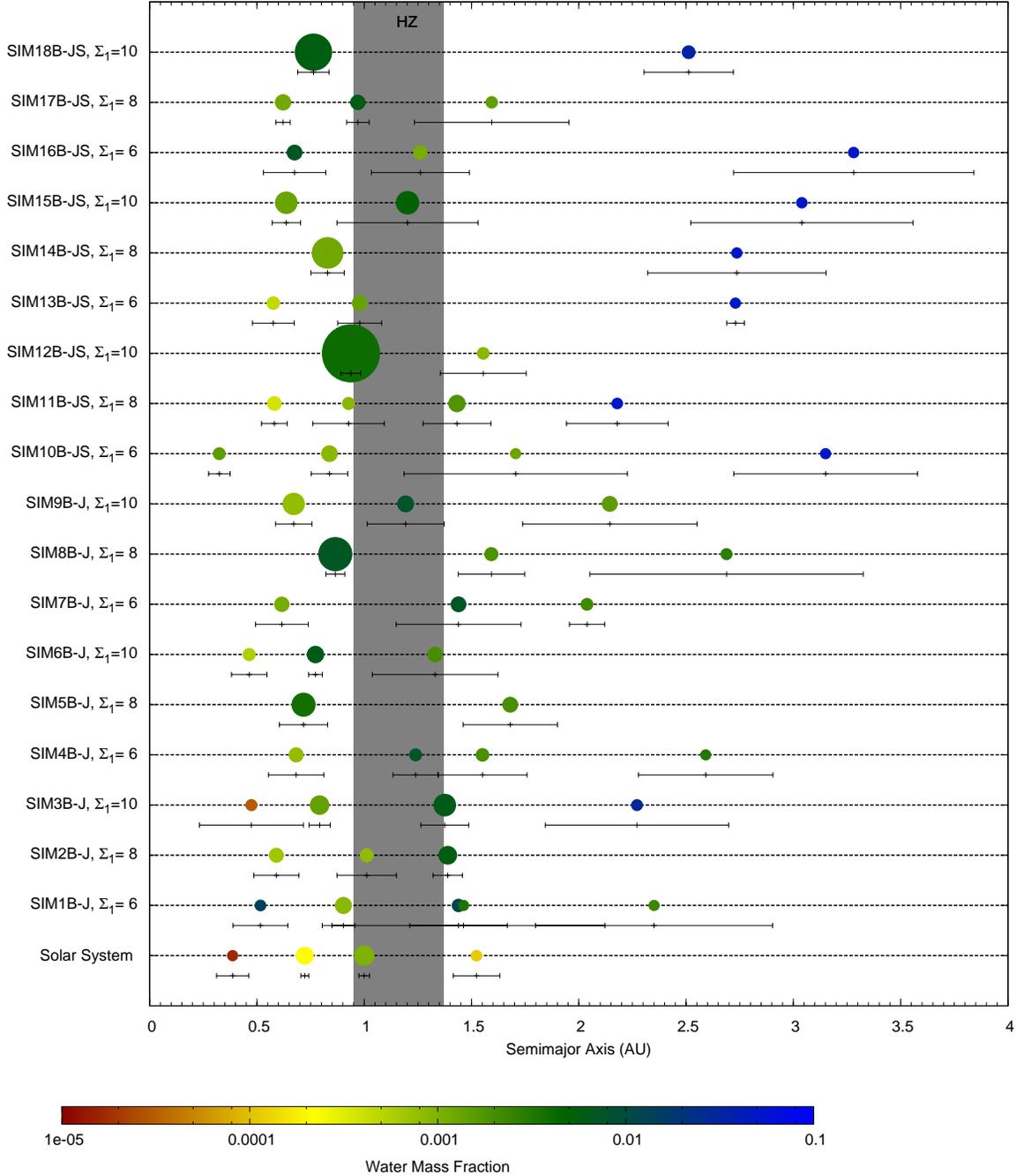}
\label{figura 1}
\caption{Final configuration of 18 simulations of model B considering only Jupiter (SIM1B-J to Sim9B-J) 
and Jupiter-Saturn (Sim10B-JS to Sim18B-JS), for different values of the disk's surface density at 1 AU.
The solar system is shown for a comparison. The size of each body corresponds to its relative physical size,
however, it is not to scale on the $x$-axis. The color of each planet represents its water-mass fraction. 
The eccentricity of each body is represented by a horizontal line depicting its variation in its heliocentric 
distance.}
\end{minipage}
\end{figure}

\clearpage
\begin{figure}
\vskip 1in
\centering
\hbox{
\includegraphics[scale=.68]{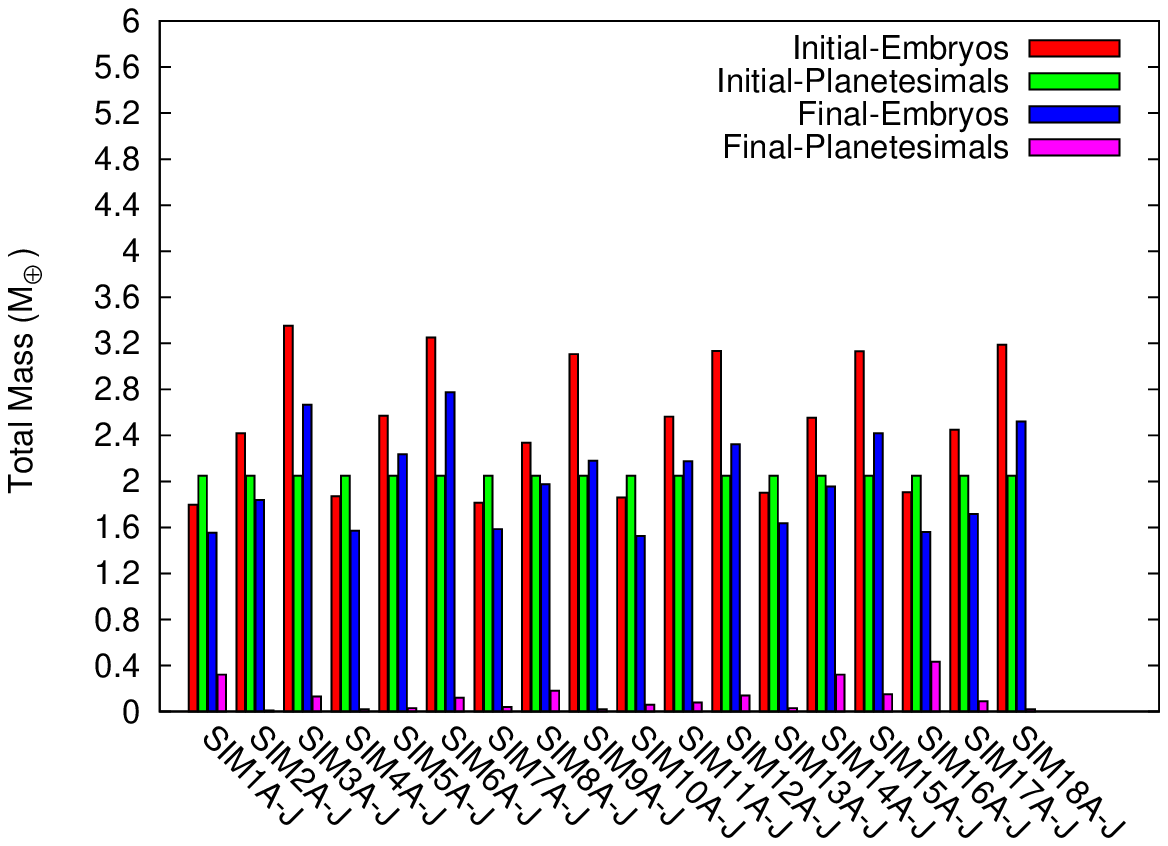}
\includegraphics[scale=.68]{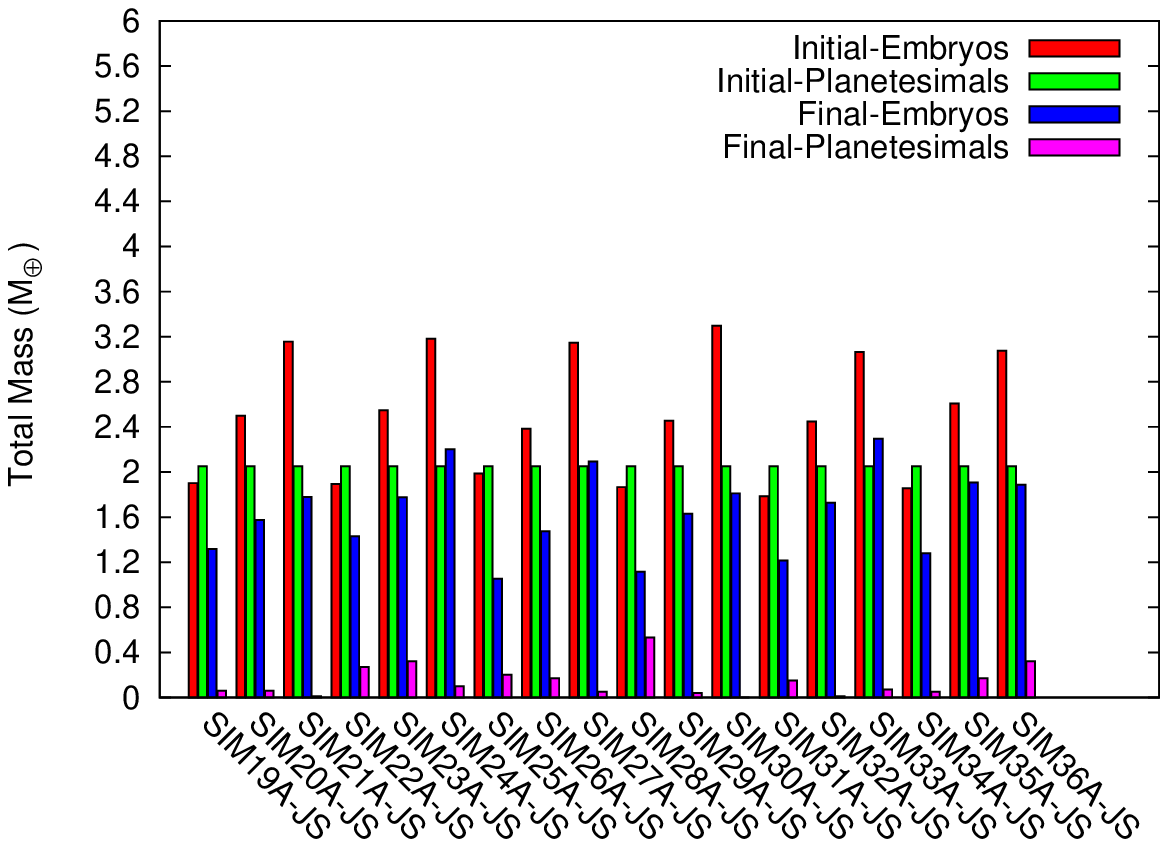}}
\hbox{
\includegraphics[scale=.68]{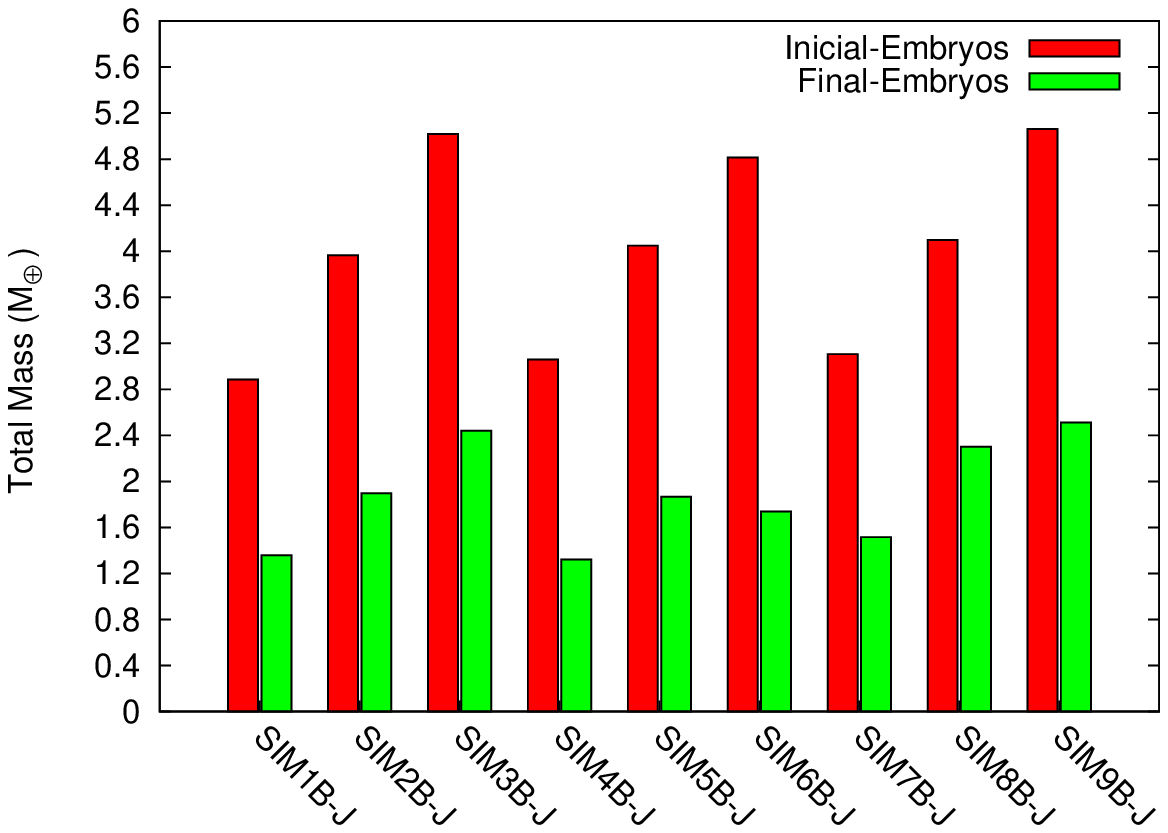}
\includegraphics[scale=.68]{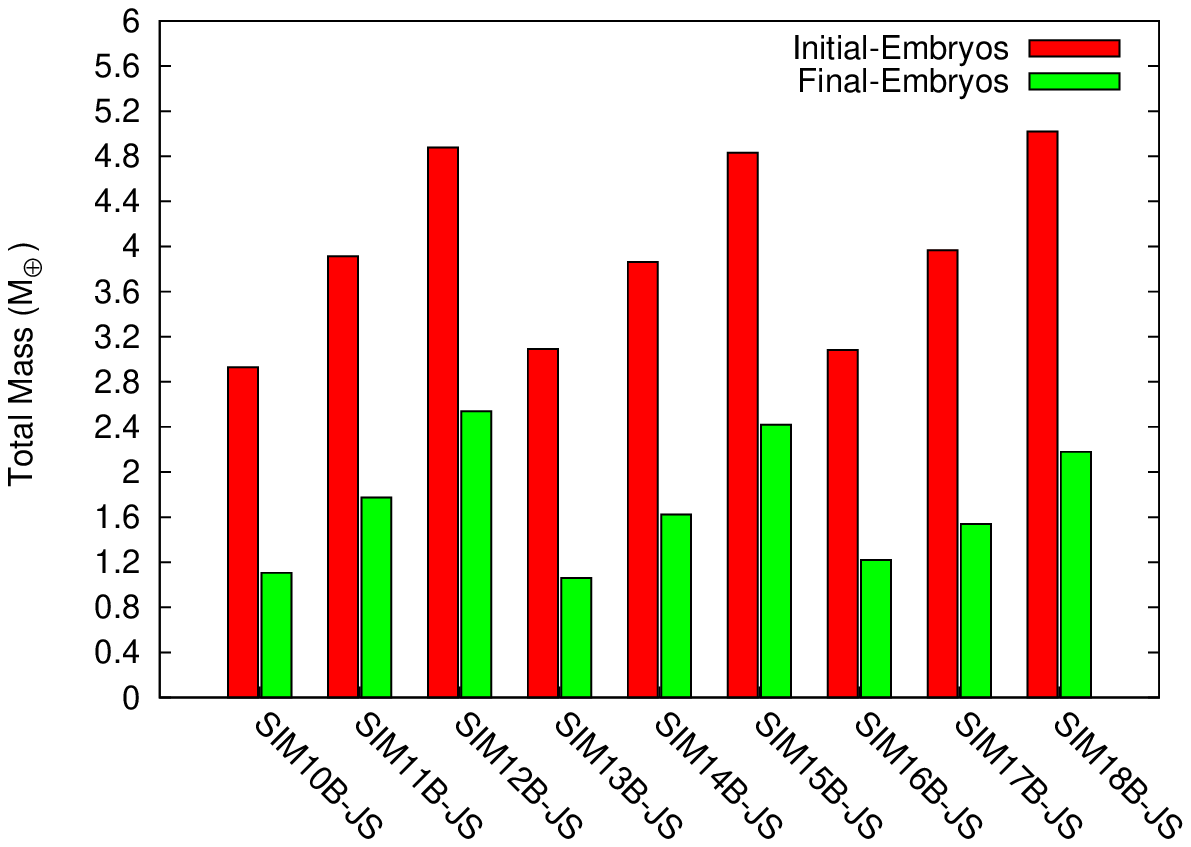}}
\caption{Initial and final  mass of embryos and planetesimals in all the simulations. Upper left: Model A with only Jupiter.
Upper right: Model A with Jupiter and Saturn. Lower left: Model B with only Jupiter. Lower right: Model B with Jupiter and Saturn}
\end{figure}

\clearpage
\begin{figure}
\vskip 0.5 in
 \begin{minipage}{170mm} 
\centering
\includegraphics[width=1\linewidth]{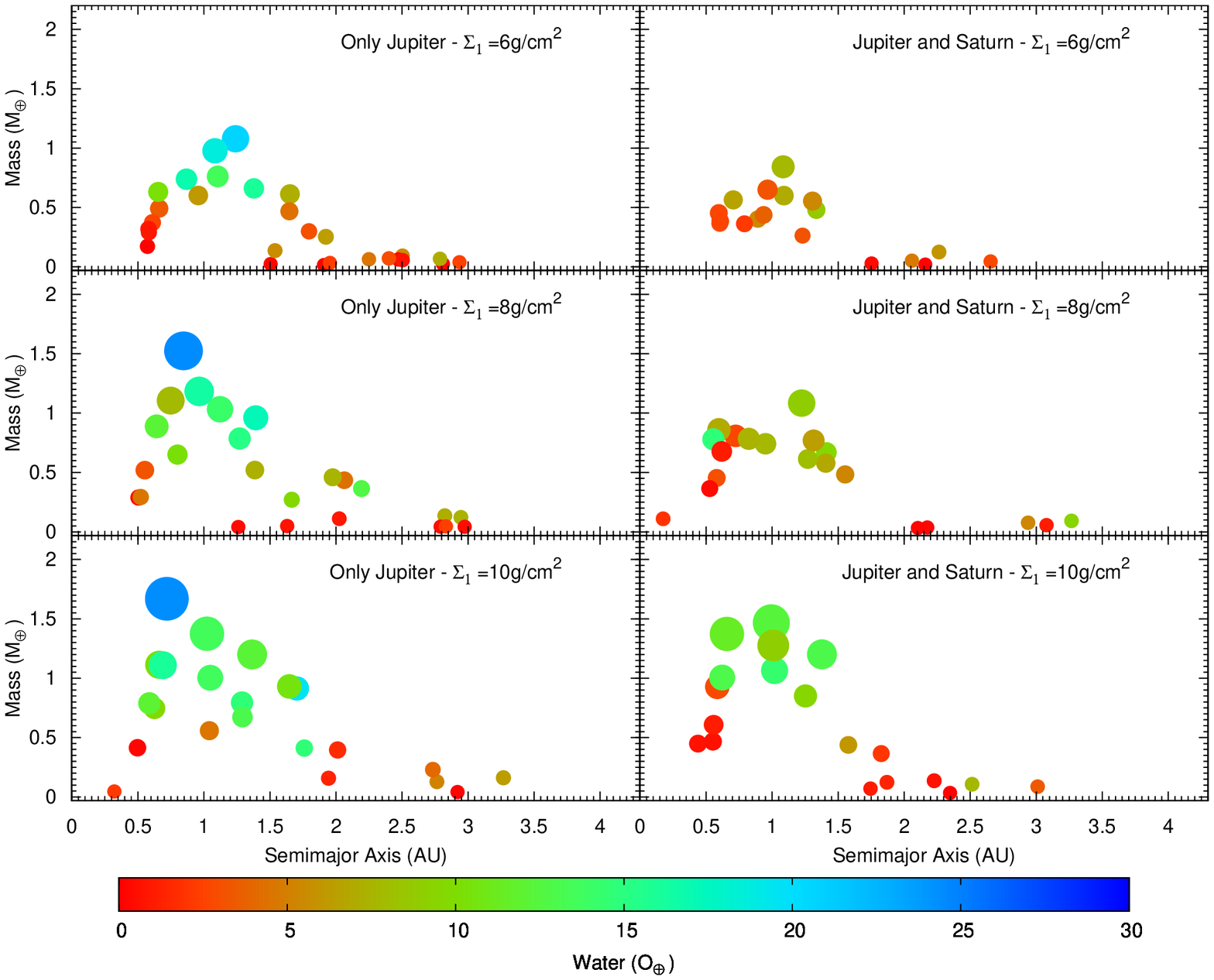}
\label{figura 1}
\caption{Final semimajor axis  versus the final mass ($M_\oplus$) of the planets in the simulations of model A. 
Note that each box shows the final results of all 6 simulations corresponding to its surface density (at 1 AU)
and giant planets configurations as explained in the text. The size of each body corresponds to its relative physical 
size, however, it is not to scale on the $x$-axis. The color of each planet represents its water content.}
\end{minipage}
\end{figure}

\clearpage
\begin{figure}
\vskip 0.5 in
 \begin{minipage}{170mm} 
\centering
\includegraphics[width=1\linewidth]{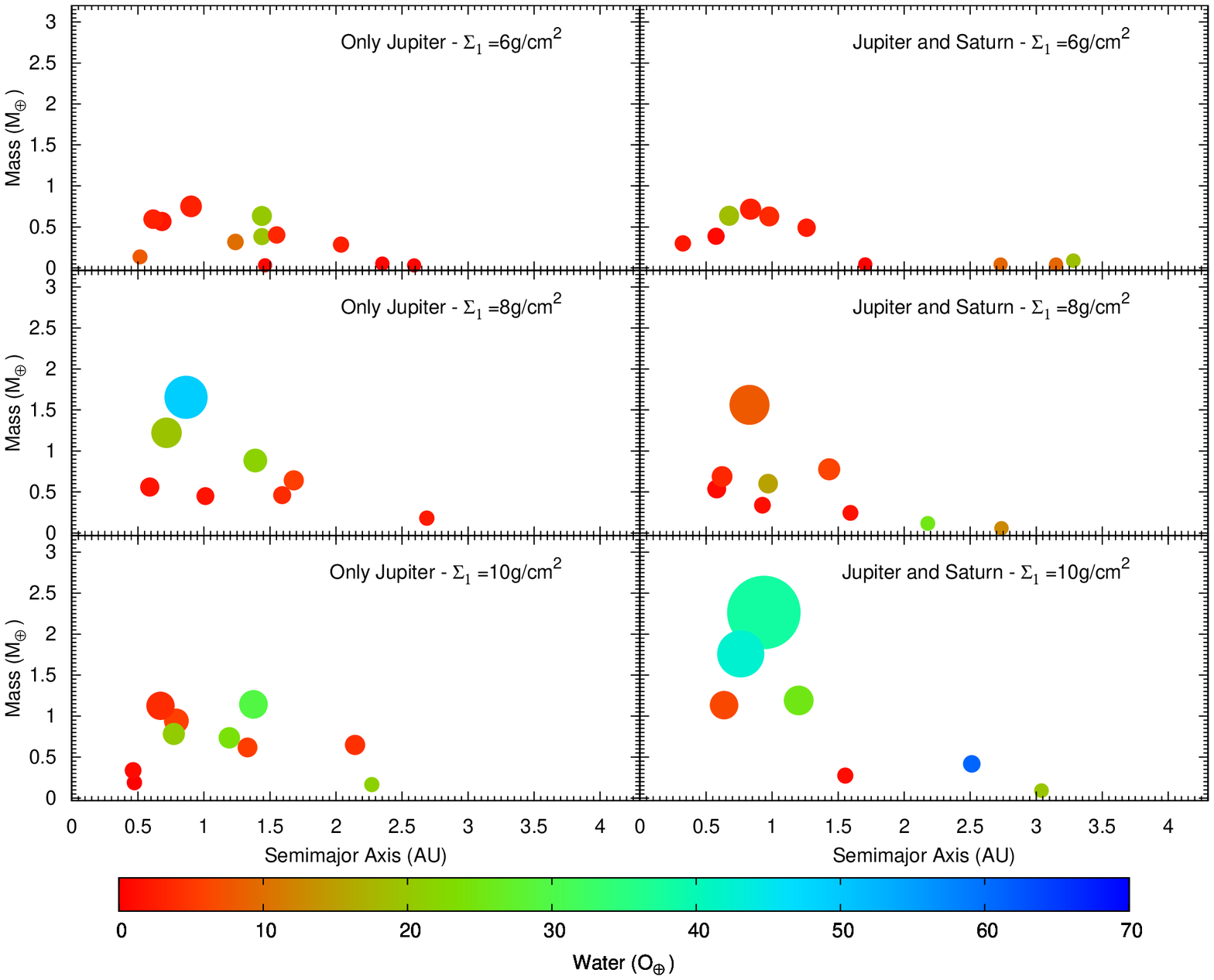}
\label{figura 1}
\caption{Final semimajor axis  versus the final mass ($M_\oplus$) of the planets in the simulations of model B. 
Note that each box shows the final results of all 3 simulations corresponding to its surface density (at 1 AU)
and giant planets configurations as explained in the text. The size of each body corresponds to its relative physical 
size, however, it is not to scale on the $x$-axis. The color of each planet represents its water content.}
\end{minipage}
\end{figure}

\clearpage
\begin{table}
\vskip 2in
\caption{Values of the D/H ratio for comets, nebular gas, and SMOW.}
   \centering 
\begin{tabular}{@{}lll@{}}
  \hline
 Body & D/H Ratio & Reference  \\
  \hline\hline
Halley      & 3.16$\pm 0.34\times 10^{-4}$    & Eberhardt et al. (1995)    \\
\hline
Hyakutake          &2.9$\pm 1.0\times 10^{-4}$   & Bockelee-Morvan et al. (1998)     \\
\hline
Hale-Bopp         &  3.3$\pm 0.8\times 10^{-4}$     & Meier et al. (1998)   \\
\hline
C/2001 Q4        &  4.6$\pm 1.4\times 10^{-4}$ & Weaver et al. (2008)     \\
\hline
C/2002 T7      &  2.5$\pm 0.7\times 10^{-4}$ & Hutsemékers (2008)     \\
\hline
8P/Tuttle       &  4.09$\pm 1.45\times 10^{-4}$ & Villanueva et al. (2009)     \\
\hline
153P/Ikeya-Zhang       &  $< 2.8\pm 0.3\times 10^{-4}$ & Biver et al. (2006)     \\
\hline
C/2004 Q2       &  $< 2.2\times 10^{-4}$ & Biver et al. (2005)     \\
\hline
Hartley 2 &         $1.61 \pm 0.24\times 10^{-4}$  &  Hartog et al. (2011)  \\
\hline
Nebula       &  2.1$\pm 0.4\times 10^{-5}$ & Lellouch et al. (2001)     \\
\hline
SMOW            & 1.49$\pm 0.03\times 10^{-4}$  & Lecuyer, Gillet \& Robert (1998) \\
\hline
\end{tabular}
\end{table}

\clearpage

\begin{table}
\vspace{-1.0cm}
 \scriptsize
\caption{Initial total mass and the amount of mass lost due to different mechanisms in simulations of model A
(J indicates only Jupiter and JS represents Jupiter-Saturn configuration). 
From left to right, the columns show the simulation's number, surface density at 1 AU, initial total mass of the disk, 
total ejected mass from the disk, the amount of the mass collided with Sun, the amount of the mass collided
with Jupiter, and the amount of the mass collided with Saturn. The unit of mass is $M_\oplus$.}
   \centering 
\begin{tabular}{@{}lccccccc@{}}
  \hline
  \\
  Sim & $\Sigma_1$      & Total     & Ejected        & Collisions with    &   Collisions with    &  Collisions with                \\
      &                  & Mass       &  Mass         &     Sun            &   Jupiter           &      Saturn                                 \\
 \hline\hline
1A-J  &    6    &   3.8480  &   1.7684   & 0.0453  &   0.1778  &      -        \\ 
\hline
2A-J  &    8   &   4.4677  &   2.2501   & 0.0200  &   0.2158  &       -       \\ 
\hline                                                                  
3A-J  &    10  &   5.4021  &   2.2587   & 0.0100  &   0.2922  &        -       \\ 
\hline                                                                    
4A-J  &   6       &   3.9221  &   2.0949   & 0.0309  &   0.1300  &     -          \\ 
\hline      
5A-J  &   8         &   4.6211  &   2.1437   & 0.0759  &   0.2000  &  -             \\ 
\hline                                                                   
6A-J  &   10        &   5.3003  &   2.1920   & 0.0681  &   0.2200  &  -             \\ 
\hline                                                                   
7A-J  &   6         &   3.8642  &   1.8874   & 0.0300  &   0.2804  &   -            \\ 
\hline                                                                   
8A-J   &  8        &   4.3866  &   1.9663   & 0.0000  &   0.2464  &    -           \\ 
\hline      
9A-J   &  10       &   5.1552  &   2.7811   & 0.0200  &   0.1800  &     -          \\ 
\hline                                                                     
10A-J   &  6        &   3.9105  &   1.9366   & 0.0100  &   0.2333  &    -           \\ 
\hline                                                                     
11A-J  &   8       &   4.6122  &   2.1914   & 0.0100  &   0.1970  &      -         \\ 
\hline                                                                     
12A-J   &  10       &   5.1824  &   2.3334   & 0.0100  &   0.2600  &     -          \\ 
\hline
13A-J  &   6       &   3.9518  &   2.0945   & 0.0100  &   0.1300  &     -          \\ 
\hline                                                                     
14A-J   &  8        &   4.6020  &   2.0644   & 0.0200  &   0.2382  &    -           \\ 
\hline                                                                     
15A-J   &  10       &   5.1811  &   2.4240   & 0.0200  &   0.1500  &     -          \\ 
\hline                                                                     
16A-J   &  6        &   3.9563  &   1.7749   & 0.0321  &   0.1700  &     -          \\ 
\hline       
17A-J   &  8        &   4.5001  &   2.3623   & 0.0000  &   0.1700  &    -           \\ 
\hline                                                                     
18A-J &    10     &   5.2367  &   2.2752   & 0.0100  &   0.3377  &      -         \\ 
\hline                                                                     
19A-JS  &  6        &   3.9516  &   2.4910   & 0.0300  &   0.0200  &      0.00                             \\ 
\hline                                                                     
20A-JS  &  8        &   4.5495  &   2.5844   & 0.1576  &   0.0652  &      0.01                             \\ 
\hline       
21A-JS  &  10       &   5.2066  &   2.8992   & 0.0980  &   0.0900  &      0.00                             \\ 
\hline
22A-JS  &  6        &   3.9444  &   2.1014   & 0.0426  &   0.0500  &      0.01                             \\ 
\hline       
23A-JS  &  8        &   4.5990  &   2.5392   & 0.0100  &   0.0300  &      0.00                             \\ 
\hline       
24A-JS  &  10       &   5.2309  &   2.9351   & 0.0100  &   0.0600  &      0.00                             \\ 
\hline
25A-JS  &  6        &   4.0377  &   2.3283   & 0.0982  &   0.0600  &      0.01                             \\ 
\hline       
26A-JS  &  8        &   4.4341  &   2.4578   & 0.0400  &   0.0700  &      0.00                             \\ 
\hline       
27A-JS  &  10       &   5.1971  &   2.7734   & 0.1220  &   0.0900  &      0.00                             \\ 
\hline
28A-JS  &  6        &   3.9163  &   1.9476   & 0.0100  &   0.0721  &      0.00                             \\ 
\hline       
29A-JS &   8       &   4.5052  &   2.5096   & 0.0200  &   0.1654  &      0.00                             \\ 
\hline       
30A-JS  &  10       &   5.3477  &   3.0595   & 0.0200  &   0.0700  &      0.02                             \\ 
\hline
31A-JS  &  6        &   3.8367  &   2.2518   & 0.1190  &   0.0400  &      0.00                             \\ 
\hline       
32A-JS  &  8        &   4.4984  &   2.6605   & 0.1018  &   0.0500  &      0.00                             \\ 
\hline       
33A-JS  &  10       &   5.1142  &   2.6800   & 0.0757  &   0.1000  &      0.00                             \\ 
\hline
34A-JS  &  6        &   3.9050  &   2.4512   & 0.0839  &   0.0516  &      0.01                             \\ 
\hline       
35A-JS  &  8        &   4.6589  &   2.3875   & 0.0200  &   0.0600  &      0.00                             \\ 
\hline       
36A-JS  &  10       &   5.1265  &   2.4865   & 0.2034  &   0.0700  &      0.00                             \\ 
\hline\hline
\end{tabular}
\end{table}

\clearpage
\begin{table}
\vskip 1in
 \scriptsize
\caption{Initial total mass and the amount of mass lost due to different mechanisms in simulations of model B
(J indicates only Jupiter and JS represents Jupiter-Saturn configuration). 
From left to right, the columns show the simulation's number, surface density at 1 AU, initial total mass of the disk, 
total ejected mass from the disk, the amount of the mass collided with Sun, the amount of the mass collided
with Jupiter, and the amount of the mass collided with Saturn. The unit of mass is $M_\oplus$.}
   \centering 
\begin{tabular}{@{}lccccccc@{}}
  \hline
  \\
  Sim & $\Sigma_1$      & Total     & Ejected        & Collisions with    &   Collisions with    &  Collisions with                \\
      &                  & Mass       &  Mass         &     Sun            &   Jupiter           &      Saturn                                 \\
 \hline\hline
1B-J  &   6     &   2.8853  &   1.3513   & 0.0000  &   0.0918  &      - \\
\hline
2B-J    &  8      &   3.9657  &   1.8604   & 0.0000  &   0.0627  &       - \\
\hline
3B-J    &   10     &   5.0178  &   1.9455   & 0.1388  &   0.2843  &    - \\
\hline
4B-J    &   6        &   3.0589  &   1.4883   & 0.0723  &   0.0356  &  - \\
\hline          
5B-J    &    8       &   4.0483  &   2.1117   & 0.0000  &   0.0000  &     - \\
\hline          
6B-J    &     10     &   4.8143  &   2.8193   & 0.0000  &   0.0650  &     - \\
\hline
7B-J    &   6        &   3.1053  &   1.3486   & 0.0740  &   0.1417  &     - \\
\hline          
8B-J    &    8       &   4.0984  &   1.5419   & 0.0282  &   0.1470  &      - \\
\hline          
9B-J    &     10     &   5.0616  &   2.4504   & 0.0000  &   0.1088  &      - \\
\hline
10B-JS   &   6        &   2.9281  &   1.6547   & 0.0000  &   0.0487  &      0.00  \\ 
\hline          
11B-JS   &    8       &   3.9137  &   1.9182   & 0.1108  &   0.0000  &      0.00  \\ 
\hline          
12B-JS   &     10     &   4.8776  &   2.4426   & 0.0000  &   0.0000  &      0.00  \\ 
\hline
13B-JS   &   6        &   3.0913  &   1.7351   & 0.0000  &   0.0000  &      0.00  \\ 
\hline          
14B-JS   &    8       &   3.8632  &   1.9304   & 0.0856  &   0.0801  &      0.00  \\ 
\hline          
15B-JS   &     10     &   4.8309  &   2.0613   & 0.0876  &   0.1550  &      0.00  \\ 
\hline
16B-JS   &   6        &   3.0828  &   1.5144   & 0.0998  &   0.0000  &      0.00  \\ 
\hline          
17B-JS   &    8       &   3.9678  &   2.1129   & 0.1262  &   0.0910  &      0.00  \\ 
\hline          
18B-JS   &     10     &   5.0213  &   2.8594   & 0.0000  &   0.0000  &      0.11  \\ 
\hline\hline
\end{tabular}
\end{table}

\clearpage

\begin{table}
 \scriptsize
\caption{Final planets inside of HZ in simulations of model A considering only Jupiter. From left to right, the columns show simulation's number, semimajor axis, eccentricity, mass ($M_\oplus$), amount of water ($O_\oplus$), percentage of asteroidal mass ($> 2.5 AU$), percentage of asteroidal water ($> 2.5 AU$), percentage of cometary water needed to raise the D/H ratio to the current value of SMOW, time (Myr) of delivery of $1 O_\oplus$, $2 O_\oplus$, $5 O_\oplus$, $10 O_\oplus$ and $15 O_\oplus$. For a comparison the values between Parentheses were obtained using the model of water distribution as in Raymond et al. (2004; 2006; 2009). When the values obtained using our model are equal to those obtained using  Raymond’s model, only one entry is shown.}
   \centering 
    \setlength{\tabcolsep}{3pt}
    \renewcommand{\arraystretch}{1.3}
\begin{tabular}{@{}lccccccccccccc@{}}
  \hline
  \\
Sim       & $a{\rm _f}$      & $e{\rm _f}$        & Mass    & Water   & \%$M_{\rm ast}$                  &  \%$W_{\rm ast}$ &   \%$W_{\rm C}$  & $t_{\rm del}$    &  $t_{\rm del}$   &  $t_{\rm del}$   &   $t_{\rm del}$ &  $t_{\rm del}$ \\
     &                           &              &            $(M_\oplus)$         &      $(O_\oplus)$                  &                       &            &                       & $1 O_{\oplus}$  & $2 O_\oplus$ & $5 O_\oplus$   & $ 10 O_\oplus$ &$ 15 O_\oplus$\\
\hline\hline
1A-J     &    1.11   &  0.11  &    0.76 &   13.5 (9.0) &    5.3    &    63.1 (95.1)  &     1.6 (-)    &        7.5 (20.8)       &        12.2 (20.8)       &       20.8 (221.4)   &      221.4 ($\times$)&        $\times$ \\ 
\hline
2A-J       &    0.96   &  0.08  &    1.18 &   16.4 (9.4) &    3.4    &    52.1 (90.5)  &    13.9 (-)    &       12.0 (20.2)       &        20.2        &       20.8 (21.4)   &       37.6 ($\times$)&      141.7 ($\times$) \\ 
\hline
3A-J      &    1.05   &  0.09  &    1.00 &   13.7 (7.0) &    3.0    &    46.8 (91.2)  &    18.8 (-)    &        8.2 (37.9)       &         8.7 (37.9)       &       37.9 (110.7)   &       68.4 ($\times$)&        $\times$ \\ 
\hline
4A-J       &    1.24   &  0.01  &    1.08 &   20.6 (13.2) &    5.6    &    62.1 (96.6)  &     2.8 (-)    &       13.3 (29.6)       &        20.3 (29.6)       &       35.0 (232.2)   &      232.2 &      232.2 ($\times$) \\ 
\hline
5A-J      &    1.27   &  0.09  &    0.79 &   15.2 (7.5) &    3.8    &    42.1 (85.1)  &    22.7 (-)    &        0.2 (10.2)       &         0.5 (10.2)       &        3.3 (68.9)   &       68.9 ($\times$)&      141.8 ($\times$) \\ 
\hline
9A-J     &    1.02   &  0.04  &    1.38 &   13.5 (5.5) &    1.5    &    31.6 (77.6)  &    30.2 (-)    &        4.6 (13.2)       &         7.4 (13.2)       &       13.2 (23.0)   &       19.7 ($\times$)&        $\times$ \\ 
\hline
12A-J       &    1.29   &  0.11  &    0.80 &   14.6 (11.3) &    6.3    &    72.9 (94.6)  &   -    &       22.5      &        22.5        &       64.8 (68.1)   &       68.1 (111.9)&         $\times$\\ 
\hline
13A-J       &    1.08   &  0.02  &    0.98 &   18.7 (13.2) &    6.2    &    68.5 (96.8)  &    -   &        2.7        &         2.7       &       37.4 (46.6)   &       46.6 (69.5)&       69.5 ($\times$) \\ 
\hline
15A-J       &    1.04   &  0.09  &    0.56 &    4.5 (2.4) &    1.8    &    47.0 (87.4)  &    18.6 (-)    &        9.1 (33.8)       &        33.8        &        $\times$   &        $\times$&        $\times$ \\ 
15A-J       &    1.29   &  0.08  &    0.67 &   13.0 (7.3) &    4.5    &    49.2 (87.3)  &    16.6 (-)    &        7.8        &         7.8        &        8.3 (67.3)   &       67.3 ($\times$)&        $\times$ \\ 
\hline
16A-J       &    0.96   &  0.04  &    0.60 &    6.1 (2.7) &    1.7    &    34.7 (78.8)  &    28.0 (-)    &       56.3 (62.1)       &        62.1        &       62.1 ($\times$)   &      $\times$&        $\times$ \\ 
\hline
17A-J      &    1.12   &  0.07  &    1.03 &   14.1 (7.1) &    2.9    &    45.4 (90.0)  &    19.9 (-)    &        0.2 (51.8)       &        12.8 (51.8)       &       51.8 (279.5)   &      242.2 ($\times$)&        $\times$ \\ 
\hline
18A-J       &    1.37   &  0.17  &    1.20 &   12.2 (7.1) &    2.5    &    52.6 (90.7)  &    13.4 (-)    &       33.9        &        33.9        &       82.2 (146.2)   &      146.2 ($\times$)&        $\times$ \\ 
\hline
\hline
\end{tabular}
\end{table}

\clearpage

\begin{table}
 \scriptsize
\caption{Final planets inside of HZ in simulations of model A considering Jupiter and Saturn. From left to right, the columns show simulation's number, semimajor axis, eccentricity, mass ($M_\oplus$), amount of water ($O_\oplus$), percentage of asteroidal mass ($> 2.5 AU$), percentage of asteroidal water ($> 2.5 AU$), percentage of cometary water needed to raise the D/H ratio to the current value of SMOW, time (Myr) of delivery of $1 O_\oplus$, $2 O_\oplus$, $5 O_\oplus$, $10 O_\oplus$ and $15 O_\oplus$. For a comparison the values between Parentheses were obtained using the model of water distribution as in Raymond et al. (2004; 2006; 2009). When the values obtained using our model are equal to those obtained using  Raymond’s model, only one entry is shown.}
   \centering 
    \setlength{\tabcolsep}{3pt}
    \renewcommand{\arraystretch}{1.3}
\begin{tabular}{@{}lccccccccccccc@{}}
  \hline
  \\
Sim       & $a{\rm _f}$      & $e{\rm _f}$        & Mass    & Water   & \%$M_{\rm ast}$                  &  \%$W_{\rm ast}$ &   \%$W_{\rm C}$  & $t_{\rm del}$    &  $t_{\rm del}$   &  $t_{\rm del}$   &   $t_{\rm del}$ &  $t_{\rm del}$ \\
     &                           &              &            $(M_\oplus)$         &      $(O_\oplus)$                  &                       &            &                       & $1 O_{\oplus}$  & $2 O_\oplus$ & $5 O_\oplus$   & $ 10 O_\oplus$ &$ 15 O_\oplus$\\
\hline\hline
19A-JS    &    1.34   &  0.11  &    0.48 &    8.7 (6.4) &    6.3    &    73.9 (100.0)  &   -    &       10.7 (75.5)       &        75.5        &       75.5 (227.3)   &        $\times$&        $\times$ \\ 
\hline
21A-JS       &    1.25   &  0.14  &    0.85 &    9.5 (4.6) &    2.4    &    44.9 (93.5)  &    20.4 (-)    &        8.7 (27.6)       &        27.6        &       52.5 ($\times$)   &        $\times$&        $\times$ \\ 
\hline
22A-JS       &    1.31   &  0.02  &    0.55 &    5.4 (2.3) &    1.8    &    39.6 (93.7)  &    24.6 (-)    &       20.1 (67.6)       &        52.7 (67.6)       &      101.5 ($\times$)   &        $\times$&        $\times$ \\ 
\hline
23A-JS        &    0.95   &  0.02  &    0.74 &    7.8 (4.3) &    2.7    &    54.9 (100.0)  &    11.0 (-)    &       18.3 (25.1)       &        25.1       &       30.2 ($\times$)   &        $\times$&        $\times$ \\ 
\hline
24A-JS        &    0.99   &  0.04  &    1.47 &   12.3 (3.1) &    0.7    &    17.3 (68.8)  &    38.3 (-)    &        6.9 (30.5)       &         9.8 (30.5)       &       16.9 ($\times$)   &       43.2 ($\times$)&        $\times$ \\ 
\hline
25A-JS      &    1.09   &  0.03  &    0.60 &    7.0 (2.4) &    1.7    &    30.3 (87.6)  &    30.9 (-)    &       12.9 (107.5)       &        48.1 (107.5)       &      107.5 ($\times$)   &        $\times$&        $\times$ \\ 
\hline
26A-JS        &    1.27   &  0.19  &    0.61 &    8.1 (4.3) &    3.3    &    53.0 (100.0)  &    13.0 (-)    &       16.1        &        16.1        &       40.6 ($\times$)   &        $\times$&        $\times$ \\ 
\hline
27A-JS       &    1.02   &  0.03  &    1.07 &   14.3 (7.0) &    2.8    &    44.8 (91.0)  &    20.4 (-)    &        0.1 (49.3)       &         0.2 (49.3)       &       40.8 (135.3)   &      135.3 ($\times$)&        $\times$ \\ 
\hline
28A-JS      &    0.94   &  0.05  &    0.44 &    3.7 (2.1) &    2.3    &    57.3 (100.0)  &     8.5 (-)    &       37.2 (63.2)       &        63.2        &        $\times$   &        $\times$&        $\times$ \\ 
\hline
29A-JS       &    1.31   &  0.11  &    0.77 &    6.4 (2.1) &    1.3    &    33.5 (100.0)  &    28.9 (-)    &       13.9 (45.2)       &        13.9 (45.2)       &       45.2 ($\times$)   &        $\times$&        $\times$ \\ 
\hline
31A-JS      &    0.97   &  0.12  &    0.65 &    3.2 (0.1) &    0.0    &     0.0  &    45.9    &       70.9 ($\times$)       &       146.4 ($\times$)       &        $\times$   &        $\times$&        $\times$ \\ 
\hline
34A-JS       &    1.08   &  0.02  &    0.84 &    7.8 (2.3) &    1.2    &    27.4 (94.0)  &    32.8 (-)    &       26.4 (98.2)       &        31.6 (98.2)       &       98.2 ($\times$)   &        $\times$&        $\times$ \\ 
\hline
35A-JS      &    1.22   &  0.05  &    1.08 &    8.9 (0.5) &    0.0    &     0.0  &    45.9    &        3.8 ($\times$)       &        35.4 ($\times$)       &       52.4 ($\times$)   &        $\times$&        $\times$ \\ 
\hline
36A-JS      &    1.01   &  0.02  &    1.28 &    9.0 (2.1) &    0.8    &    23.7 (100.0)  &    34.9 (-)    &       14.3 (17.3)       &        14.6 (17.3)       &       17.3 ($\times$)   &        $\times$&        $\times$ \\ 
\hline
\hline
\end{tabular}
\end{table}

\clearpage
\begin{table}
 \scriptsize
\caption{Final planets inside of HZ in simulations of model B (J indicates only Jupiter and JS represents Jupiter-Saturn configuration). From left to right, the columns show simulation's number, semimajor axis, eccentricity, mass ($M_\oplus$), amount of water ($O_\oplus$), percentage of asteroidal mass ($> 2.5 AU$), percentage of asteroidal water ($> 2.5 AU$), percentage of cometary water needed to raise the D/H ratio to the current value of SMOW, time (Myr) of delivery of $1 O_\oplus$, $2 O_\oplus$, $5 O_\oplus$, $10 O_\oplus$ and $15 O_\oplus$. For a comparison the values between Parentheses were obtained using the model of water distribution as in Raymond et al. (2004; 2006; 2009). When the values obtained using our model are equal to those obtained using  Raymond’s model, only one entry is shown.}
  \begin{minipage}{\linewidth} 
   \centering 
    \setlength{\tabcolsep}{3pt}
    \renewcommand{\arraystretch}{1.3}
\begin{tabular}{@{}lcccccccccccccc@{}}
  \hline
  \\
Sim       & $a{\rm _f}$      & $e{\rm _f}$        & Mass    & Water   & \%$M_{\rm ast}$                  &  \%$W_{\rm ast}$ &   \%$W_{\rm C}$  & $t_{\rm del}$    &  $t_{\rm del}$   &  $t_{\rm del}$   &   $t_{\rm del}$ &  $t_{\rm del}$ \\
     &                           &              &            $(M_\oplus)$         &      $(O_\oplus)$                  &                       &            &                       & $1 O_{\oplus}$  & $2 O_\oplus$ & $5 O_\oplus$   & $ 10 O_\oplus$ &$ 15 O_\oplus$\\
\hline\hline
1B-J    &    0.90   &  0.06  &    0.75 &    2.8 (0.1) &    0.0    &     0.0 (0.0)  &    45.9     &       33.1 ($\times$)       &        33.1 ($\times$)       &        $\times$   &        $\times$&         $\times$ \\ 
\hline
2B-J     &    1.01   &  0.14  &    0.45 &    1.6 (0.0) &    0.0    &     0.0 (0.0)  &    45.9     &       24.1 ($\times$)       &         $\times$       &         $\times$   &         $\times$&         $\times$ \\ 
\hline
4B-J     &    1.24   &  0.09  &    0.32 &    9.9 (8.2) &   11.9    &    82.0 (98.2)  &   -    &       81.5       &        81.5        &       81.5    &         $\times$&         $\times$ \\ 
\hline
6B-J      &    1.33   &  0.22  &    0.62 &    5.3 (0.9) &    0.0    &     0.0 (0.0)  &    45.9     &        0.001 ($\times$)       &         0.001 ($\times$)       &       41.4 ($\times$)   &         $\times$&         $\times$ \\ 
\hline
9B-J       &    1.19   &  0.15  &    0.74 &   24.0 (17.9) &   11.1    &    72.0 (96.5)  &   -    &        0.0 (1.5)       &         0.2 (1.5)       &        1.5   &        1.5 &        1.5  \\ 
\hline
11B-JS    &    0.93   &  0.18  &    0.34 &    1.3 (0.0) &    0.0    &     0.0   &    45.9    &       41.5 ($\times$)       &         $\times$       &        $\times$   &        $\times$&        $\times$ \\ 
\hline
12B-JS       &    0.94   &  0.05  &    2.26 &   38.1 (23.2) &    4.3    &    54.3 (89.2)  &    11.7 (-)    &        7.5 (31.2)       &        17.7 (31.2)       &       31.2    &       31.2 &       31.2  \\ 
\hline
13B-JS        &    0.98   &  0.10  &    0.63 &    3.6 (0.0) &    0.0    &     0.0  &    45.9 &       30.5 ($\times$)       &        71.6 ($\times$)       &        $\times$   &        $\times$&        $\times$ \\ 
\hline
15B-JS        &    1.20   &  0.27  &    1.19 &   25.2 (17.9) &    7.0    &    69.9 (98.1)  &    -   &        0.0 (12.3)       &         0.6 (12.3)       &       12.3    &       12.3 &       12.3  \\ 
\hline
16B-JS        &    1.26   &  0.18  &    0.49 &    2.3 (0.0) &    0.0    &     0.0 (0.0)  &    45.9     &       41.9 ($\times$)       &        61.2 ($\times$)       &        $\times$   &        $\times$&        $\times$ \\ 
\hline
17B-JS       &    0.97   &  0.05  &    0.60 &   15.4 (12.9) &    9.7    &    80.7( 96.5)  &   -  &       26.5        &        26.5        &       26.5    &       26.5 &      146.6 ($\times$) \\ 
\hline
\end{tabular}
\end{minipage}
\end{table}

\clearpage
\begin{table}
\vspace{-1.0cm}
 \scriptsize
\caption{Analysis of the amount and time of water-delivery in planets inside of HZ (see Tables 4, 5 and 6). From left to right the columns show the disk model (J indicates only Jupiter and JS represents Jupiter-Saturn configuration), water distribution model, amount of water delivered, number of planets that received such amount of water, mean water-delivery time, range of the water-delivery time, mean of the 60\% accretion time, range of the 60\% accretion time, mean-time of the last giant collision, and the range of the time of the last giant collision.
}
\begin{minipage}{\linewidth} 
   \centering 
   \renewcommand{\arraystretch}{0.7}
\begin{tabular}{@{}cccllll@{}}
  \hline
  \hline
Model  &    Water       	& Water               &  N   & Mean-time of (Myr)    &  Mean-time (Myr) & Mean-time (Myr) of Last  \\
     &  Distribution Model   & Amount              &      &  delivery (range)     &  of 60\% accretion (range) &Giant Impact  (range)\\
  \hline
  \hline
     &                 &                         &      &      &  &\\
A-J  &                 &                     & All (13)   &     &  61.2 (9.11 - 232.2) & 159.6 (48.87 - 291)  \\
     & Our Model       &                         &      &            &    &  \\
     &                 &  $\geq 2O_{\oplus}$     & 13   & 19   (0.5 - 62.1)       &       &   \\
     &                 &  $\geq 5O_{\oplus}$     & 12   & 36   (3.3 - 82.2)       &       &   \\
     &                 &  $\geq 10O_{\oplus}$    & 11   & 110 (19.7 - 242.2 )      &       &    \\
     &                 &  $\geq 15O_{\oplus}$    & 4    & 146 (69.5 - 232.2  )     &       &     \\
     \\
     & Raymond's Model &                         &      &               &  & \\
     &                 &  $\geq 2O_{\oplus}$     & 13     & 26   (2.7 - 62.1)  &     &          \\
     &                 &  $\geq 5O_{\oplus}$     & 11   & 117  (21.4 - 279.5)  &   &            \\
     &                 &  $\geq 10O_{\oplus}$    & 2   & 90.7 (69.5- 111.9 )   &     &         \\
     &                 &  $\geq 15O_{\oplus}$    & 0    &       -        &  & \\
\hline
     &                 &                         &      &        & & \\
A-JS  &                 &                     & All (14)   &     & 44.4 (16.9 - 98.18)   &145.6 (30.2 - 268.8)   \\
     & Our Model       &                         &      &        &    &      \\
     &                 &  $\geq 2O_{\oplus}$     & 14   & 40   (0.2 - 75.5 )   &     &         \\
     &                 &  $\geq 5O_{\oplus}$     & 12   & 56   (16.9 -107.5)   &     &         \\
     &                 &  $\geq 10O_{\oplus}$    & 2   & 89 (43.2 - 135.3 )    &   &          \\
     &                 &  $\geq 15O_{\oplus}$    & 0    & -              & &  \\
     \\
     & Raymond's Model &                         &      &              &  &  \\
     &                 &  $\geq 2O_{\oplus}$     & 12   & 52   (16.1 - 107.5) &     &           \\
     &                 &  $\geq 5O_{\oplus}$     & 2   & 181  (135.3 - 227.3)  &   &            \\
     &                 &  $\geq 10O_{\oplus}$    & 0   & -          &    &   \\
     &                 &  $\geq 15O_{\oplus}$    & 0    &       -   &     &  \\
\hline
     &                 &                         &      &      &    &  \\
B-J  &                 &                      & All (5)   &    & 32.2 (1.54 - 81.48)  & 56 (37.9 - 61.7)       \\
     & Our Model       &                         &      &      &      &      \\
     &                 &  $\geq 2O_{\oplus}$     & 4   & 29 (0.001 - 81.5 )    &       &      \\
     &                 &  $\geq 5O_{\oplus}$     & 3   & 41   (1.5 -81.5)      &    &       \\
     &                 &  $\geq 10O_{\oplus}$    & 1   & 1.5 (1.5 )            &  &   \\
     &                 &  $\geq 15O_{\oplus}$    & 1    & 1.5     (1.5 )       &   &  \\
     \\
     & Raymond's Model &                         &      &               &  & \\
     &                 &  $\geq 2O_{\oplus}$     & 2   & 42   (1.5 - 81.5)      &     &      \\
     &                 &  $\geq 5O_{\oplus}$     & 2   & 42   (1.5 - 81.5)      &     &      \\
     &                 &  $\geq 10O_{\oplus}$    & 1  & 1.5 (1.5 )       &     &   \\
     &                 &  $\geq 15O_{\oplus}$    & 0   & 1.5 (1.5 )      &   &    \\
\hline
     &                 &                         &      &        &  & \\
B-JS  &                 &                      & All (6)   &   &  47.17 (26.48 - 71.57)  & 105.6 (47.9 - 159.6)    \\
     & Our Model       &                         &      &        &     &     \\
     &                 &  $\geq 2O_{\oplus}$     & 5   & 36 (0.6 - 71.6 )       &   &       \\
     &                 &  $\geq 5O_{\oplus}$     & 3   & 23.3   (12.3 -31.2)    &     &        \\
     &                 &  $\geq 10O_{\oplus}$    &  3   & 23.3   (12.3 -31.2)   &       &       \\
     &                 &  $\geq 15O_{\oplus}$    & 3    & 63     (12.3 - 146.6 )      &   &         \\
     \\
     & Raymond's Model &                         &      &         &    &     \\
     &                 &  $\geq 2O_{\oplus}$     & 3   & 23.3   (12.3 -31.2)         &  &      \\
     &                 &  $\geq 5O_{\oplus}$     & 3   & 23.3   (12.3 -31.2)        &  &  \\
     &                 &  $\geq 10O_{\oplus}$    & 3   & 23.3   (12.3 -31.2)        &   &  \\
     &                 &  $\geq 15O_{\oplus}$    & 2   & 22     (21.75 )       &    &  \\
\hline

\end{tabular}
\end{minipage}
\end{table}

\clearpage
\appendix
    \setcounter{table}{0}
    \renewcommand\thetable{\Alph{section}\arabic{table}}
\section{Final configuration of the simulations  of model A}

\begin{table}
 \scriptsize
   \renewcommand{\arraystretch}{0.8}
       \setlength{\tabcolsep}{5pt}
\caption{Final results of the simulations of model A  (J indicates only Jupiter and JS represents Jupiter-Saturn configuration). From left to right, the columns show  simulation's number, surface density at 1 AU,  semimajor axis,  eccentricity, mass ($M_\oplus$),   amount of water ($O_\oplus$), percentage of asteroidal mass ($> 2.5 AU$), percentage of asteroidal water ($> 2.5 AU$), final D/H ratio, and the time (Myr) of the delivery of $1 O_\oplus$, $2 O_\oplus$, $5 O_\oplus$, $10 O_\oplus$ and $15 O_\oplus$.}
   \centering 
\begin{tabular}{@{}lcccccccccrrrr@{}}
  \hline
  \\
Sim & $\Sigma_1$        & $a_{\rm f}$      & $e_{\rm f}$        & Mass $(M_\oplus)$  & Water $(O_\oplus)$   & \%$M_{\rm ast}$                 &  \%$W_{\rm ast}$ & $D/H$    & $t_{\rm del}$    &  $t_{\rm del}$   &  $t_{\rm del}$   &   $t_{\rm del}$ &  $t_{\rm del}$ \\
     &                &            &              &                      &                        &                       &            &     & $1 O_\oplus$  & $2 O_\oplus$ & $5 O_\oplus$   & $ 10 O_\oplus$ &$ 15 O_\oplus$ \\
\hline\hline
1A-J  &  6   &    0.66   &  0.02  &    0.49    &     3.5 &       2.0 &    61.4  &   1.431e-04   &      4.8         &       4.8      &   $\times$            &   $\times$      &   $\times$   \\ 
1A-J   &   6  &    1.11   &  0.11  &    0.76    &    13.5 &       5.3 &    63.1  &   1.465e-04   &      7.5         &      12.2      &     20.8            &    221.4      &   $\times$   \\ 
1A-J    &   6  &    1.80   &  0.17  &    0.30    &     3.0 &       0.0 &     0.0  &   2.100e-05   &      5.1         &      26.4      &   $\times$            &   $\times$      &   $\times$   \\ 
\hline
2A-J   &   8  &    0.51   &  0.16  &    0.29    &     0.4 &       0.0 &     0.0  &   2.100e-05   &   $\times$         &    $\times$      &   $\times$            &   $\times$      &   $\times$   \\ 
2A-J    &   8  &    0.96   &  0.08  &    1.18    &    16.4 &       3.4 &    52.1  &   1.246e-04   &     12.0         &      20.2      &     20.8            &     37.6      &    141.7   \\ 
2A-J    &   8  &    2.19   &  0.13  &    0.37    &    12.7 &      11.0 &    67.3  &   1.550e-04   &      0.2         &       0.2      &      2.5            &    113.2      &   $\times$   \\ 
\hline
3A-J    &  10   &    0.63   &  0.04  &    0.75    &     9.6 &       5.4 &    88.6  &   1.972e-04   &     73.4         &      73.4      &     96.1            &   $\times$      &   $\times$   \\ 
3A-J    &  10   &    1.05   &  0.09  &    1.00    &    13.7 &       3.0 &    46.8  &   1.141e-04   &      8.2         &       8.7      &     37.9            &     68.4      &   $\times$   \\ 
3A-J    &  10   &    1.71   &  0.09  &    0.91    &    19.5 &       5.5 &    54.8  &   1.300e-04   &      0.0         &       1.0      &      5.7            &     12.1      &     22.4   \\ 
\hline
4A-J   &  6   &    0.61   &  0.09  &    0.37    &     2.6 &       2.7 &    82.6  &   1.854e-04   &    206.4         &     206.4      &   $\times$            &   $\times$      &   $\times$   \\ 
4A-J   &  6   &    1.24   &  0.01  &    1.08    &    20.6 &       5.6 &    62.1  &   1.446e-04   &     13.3         &      20.3      &     35.0            &    232.2      &    232.2   \\ 
4A-J   &  6   &    2.50   &  0.18  &    0.09    &     5.6 &      21.3 &    76.3  &   1.728e-04   &      1.4         &       2.2      &      2.6            &   $\times$      &   $\times$   \\ 
4A-J   &  6   &    2.81   &  0.16  &    0.03    &     0.2 &       0.0 &     0.0  &   2.100e-05   &   $\times$         &    $\times$      &   $\times$            &   $\times$      &   $\times$   \\ 
\hline
5A-J   &   8  &    0.64   &  0.08  &    0.89    &    12.1 &       4.5 &    70.3  &   1.609e-04   &     10.8         &      10.8      &     34.4            &     84.6      &   $\times$   \\ 
5A-J   &   8  &    1.27   &  0.09  &    0.79    &    15.2 &       3.8 &    42.1  &   1.047e-04   &      0.2         &       0.5      &      3.3            &     68.9      &    141.8   \\ 
5A-J   &   8  &    2.06   &  0.07  &    0.44    &     4.4 &       2.3 &    48.6  &   1.176e-04   &      5.0         &       5.0      &   $\times$            &   $\times$      &   $\times$   \\ 
5A-J   &   8  &    2.95   &  0.23  &    0.12    &     7.5 &      24.4 &    85.7  &   1.915e-04   &      0.0         &       0.0      &      0.8            &   $\times$      &   $\times$   \\ 
\hline
6A-J    &    10 &    0.72   &  0.03  &    1.67    &    24.4 &       3.6 &    52.5  &   1.256e-04   &      4.8         &       9.9      &      9.9            &      9.9      &     16.9   \\ 
6A-J    &    10 &    1.65   &  0.11  &    0.93    &    10.6 &       2.2 &    40.4  &   1.015e-04   &      6.3         &       6.3      &     12.3            &     54.8      &   $\times$   \\ 
6A-J    &   10  &    2.76   &  0.16  &    0.13    &     5.1 &      15.5 &    82.9  &   1.859e-04   &      5.4         &       5.4      &     54.1            &   $\times$      &   $\times$   \\ 
6A-J    &   10  &    2.92   &  0.15  &    0.04    &     0.2 &       0.0 &     0.0  &   2.100e-05   &   $\times$         &    $\times$      &   $\times$            &   $\times$      &   $\times$   \\ 
\hline
7A-J    &  6   &    0.65   &  0.01  &    0.63    &    10.2 &       6.4 &    83.5  &   1.872e-04   &     16.0         &      16.0      &    125.7            &    272.4      &   $\times$   \\ 
7A-J    & 6    &    1.38   &  0.03  &    0.66    &    16.0 &       7.6 &    66.7  &   1.537e-04   &      4.6         &       4.6      &     31.9            &    101.5      &    230.5   \\ 
7A-J    & 6    &    2.25   &  0.19  &    0.06    &     4.5 &      31.0 &    93.8  &   2.077e-04   &      3.5         &       3.5      &   $\times$            &   $\times$      &   $\times$   \\ 
7A-J    & 6    &    2.47   &  0.34  &    0.06    &     0.3 &       0.0 &     0.0  &   2.100e-05   &   $\times$         &    $\times$      &   $\times$            &   $\times$      &   $\times$   \\ 
7A-J    & 6    &    2.51   &  0.08  &    0.06    &     0.7 &       0.0 &     0.0  &   2.100e-05   &   $\times$         &    $\times$      &   $\times$            &   $\times$      &   $\times$   \\ 
7A-J    & 6    &    2.79   &  0.12  &    0.07    &     7.1 &      44.4 &    90.1  &   2.003e-04   &      2.5         &       2.5      &     17.0            &   $\times$      &   $\times$   \\ 
7A-J    &  6   &    2.94   &  0.17  &    0.04    &     2.5 &      25.6 &    86.8  &   1.937e-04   &     10.9         &      10.9      &   $\times$            &   $\times$      &   $\times$   \\ 
\hline
8A-J    &  8   &    0.85   &  0.05  &    1.52    &    24.4 &       5.3 &    69.8  &   1.600e-04   &      0.0         &      12.1      &     38.5            &    112.8      &    150.5   \\ 
8A-J    &  8   &    1.67   &  0.08  &    0.27    &     9.7 &      11.1 &    66.2  &   1.527e-04   &      0.4         &       0.4      &    149.5            &   $\times$      &   $\times$   \\ 
8A-J    &  8   &    2.82   &  0.13  &    0.14    &     7.1 &      22.0 &    89.7  &   1.996e-04   &    140.9         &     140.9      &    197.0            &   $\times$      &   $\times$   \\ 
8A-J    &  8   &    2.97   &  0.32  &    0.04    &     0.5 &       0.0 &     0.0  &   2.100e-05   &   $\times$         &    $\times$      &   $\times$            &   $\times$      &   $\times$   \\ 
\hline
9A-J    &  10   &    0.50   &  0.25  &    0.42    &     0.2 &       0.0 &     0.0  &   2.100e-05   &   $\times$         &    $\times$      &   $\times$            &   $\times$      &   $\times$   \\ 
9A-J    &   10  &    1.02   &  0.04  &    1.38    &  13.5 &       1.5 &    31.6  &   8.382e-05   &      4.6         &       7.4      &     13.2            &     19.7      &   $\times$   \\ 
9A-J    &   10  &    1.94   &  0.21  &    0.16    &     1.3 &       0.0 &     0.0  &   2.100e-05   &     37.9         &    $\times$      &   $\times$            &   $\times$      &   $\times$   \\ 
9A-J    &  10   &    2.73   &  0.05  &    0.23    &     4.0 &       4.3 &    53.3  &   1.270e-04   &      0.1         &       4.4      &   $\times$            &   $\times$      &   $\times$   \\ 
\hline
10A-J    &  6   &    0.57   &  0.28  &    0.17    &     0.1 &       0.0 &     0.0  &   2.100e-05   &   $\times$         &    $\times$      &   $\times$            &   $\times$      &   $\times$   \\ 
10A-J    &   6  &    0.87   &  0.02  &    0.74    &    16.8 &       8.2 &    76.2  &   1.725e-04   &     11.8         &      11.8      &     33.0            &     77.1      &    214.1   \\ 
10A-J    &  6   &    1.65   &  0.02  &    0.61    &     7.2 &       1.6 &    29.6  &   8.000e-05   &     14.4         &      28.4      &     71.4            &   $\times$      &   $\times$   \\ 
\hline
11A-J   &  8   &    0.75   &  0.02  &    1.11    &     7.8 &       0.9 &    27.4  &   7.544e-05   &      6.2         &       6.2      &     58.8            &   $\times$      &   $\times$   \\ 
11A-J   &   8  &    1.26   &  0.29  &    0.04    &     0.4 &       0.0 &     0.0  &   2.100e-05   &   $\times$         &    $\times$      &   $\times$            &   $\times$      &   $\times$   \\ 
11A-J   & 8    &    1.39   &  0.11  &    0.52    &     7.3 &       3.9 &    58.8  &   1.380e-04   &     11.7         &      11.7      &     17.4            &   $\times$      &   $\times$   \\ 
11A-J   &  8   &    1.98   &  0.11  &    0.46    &     7.5 &       2.2 &    28.4  &   7.759e-05   &      0.3         &       7.4      &      7.4            &   $\times$      &   $\times$   \\ 
11A-J   &  8   &    2.79   &  0.09  &    0.05    &     0.5 &       0.0 &     0.0  &   2.100e-05   &   $\times$         &    $\times$      &   $\times$            &   $\times$      &   $\times$   \\ 
\hline

\end{tabular}
\end{table}

\begin{table}
\scriptsize
\renewcommand{\arraystretch}{0.8}
\setlength{\tabcolsep}{5pt}
\captcont{Final results of the simulations of model A  (J indicates only Jupiter and JS represents Jupiter-Saturn configuration). From left to right, the columns show  simulation's number, surface density at 1 AU,  semimajor axis,  eccentricity, mass ($M_\oplus$),   amount of water ($O_\oplus$), percentage of asteroidal mass ($> 2.5 AU$), percentage of asteroidal water ($> 2.5 AU$), final D/H ratio, and the time (Myr) of the delivery of $1 O_\oplus$, $2 O_\oplus$, $5 O_\oplus$, $10 O_\oplus$ and $15 O_\oplus$.}
   \centering 
\begin{tabular}{@{}lcccccccccrrrr@{}}
  \hline
  \\
Sim & $\Sigma_1$        & $a_{\rm f}$      & $e_{\rm f}$        & Mass $(M_\oplus)$  & Water $(O_\oplus)$   & \%$M_{\rm ast}$                 &  \%$W_{\rm ast}$ & $D/H$    & $t_{\rm del}$    &  $t_{\rm del}$   &  $t_{\rm del}$   &   $t_{\rm del}$ &  $t_{\rm del}$ \\
     &                &            &              &                      &                        &                       &            &     & $1 O_\oplus$  & $2 O_\oplus$ & $5 O_\oplus$   & $ 10 O_\oplus$ &$ 15 O_\oplus$ \\
\hline\hline
12A-J    &   10  &    0.66   &  0.08  &    1.11    &    10.0 &       1.8 &    42.6  &   1.058e-04   &      9.6         &       9.6      &     50.8            &    121.1      &   $\times$   \\ 
12A-J    &  10   &    1.29   &  0.11  &    0.80    &    14.6 &       6.3 &    72.9  &   1.661e-04   &     22.5         &      22.5      &     64.8            &     68.1      &   $\times$   \\ 
12A-J    & 10   &    1.76   &  0.12  &    0.41    &    14.6 &      12.2 &    73.0  &   1.662e-04   &      0.0         &       0.2      &      7.2            &     58.6      &   $\times$   \\ 
\hline
13A-J    &  6   &    0.58   &  0.16  &    0.29    &     0.8 &       0.0 &     0.0  &   2.100e-05   &   $\times$         &    $\times$      &   $\times$            &   $\times$      &   $\times$   \\ 
13A-J    &   6  &    1.08   &  0.02  &    0.98    &    18.7 &       6.2 &    68.5  &   1.573e-04   &      2.7         &       2.7      &     37.4            &     46.6      &     69.5   \\ 
13A-J    & 6    &    1.51   &  0.22  &    0.02    &     0.2 &       0.0 &     0.0  &   2.100e-05   &   $\times$         &    $\times$      &   $\times$            &   $\times$      &   $\times$   \\ 
13A-J    &  6   &    1.91   &  0.37  &    0.02    &     0.0 &       0.0 &     0.0  &   2.100e-05   &   $\times$         &    $\times$      &   $\times$            &   $\times$      &   $\times$   \\ 
13A-J    &  6   &    1.92   &  0.09  &    0.25    &     6.4 &       7.9 &    67.0  &   1.543e-04   &     20.0         &      20.0      &     55.5            &   $\times$      &   $\times$   \\ 
13A-J    &  6   &    2.40   &  0.06  &    0.07    &     2.9 &      13.9 &    73.0  &   1.663e-04   &     18.0         &      18.0      &   $\times$            &   $\times$      &   $\times$   \\ 
\hline
14A-J    &   8  &    0.52   &  0.14  &    0.29    &     4.6 &       6.8 &    92.4  &   2.049e-04   &    126.1         &     126.1      &   $\times$            &   $\times$      &   $\times$   \\ 
14A-J    & 8    &    0.80   &  0.05  &    0.65    &    10.3 &       6.2 &    83.0  &   1.861e-04   &     28.4         &      28.4      &     44.8            &    212.2      &   $\times$   \\ 
14A-J    &  8   &    1.39   &  0.10  &    0.96    &    17.2 &       4.2 &    49.6  &   1.197e-04   &     11.4         &      14.3      &     40.8            &     40.8      &    151.5   \\ 
14A-J    &  8   &    2.83   &  0.21  &    0.05    &     2.4 &      21.1 &    88.7  &   1.975e-04   &      0.4         &       0.4      &   $\times$            &   $\times$      &   $\times$   \\ 
\hline
15A-J   &  10   &    0.59   &  0.02  &    0.79    &    12.0 &       5.1 &    71.1  &   1.625e-04   &     14.1         &      14.1      &     30.3            &    110.1      &   $\times$   \\ 
15A-J   &  10   &    1.04   &  0.09  &    0.56    &     4.5 &       1.8 &    47.0  &   1.146e-04   &      9.1         &      33.8      &   $\times$            &   $\times$      &   $\times$   \\ 
15A-J   &   10  &    1.29   &  0.08  &    0.67    &    13.0 &       4.5 &    49.2  &   1.189e-04   &      7.8         &       7.8      &      8.3            &     67.3      &   $\times$   \\ 
15A-J   &  10   &    2.01   &  0.09  &    0.40    &     1.6 &       0.0 &     0.0  &   2.100e-05   &     21.7         &    $\times$      &   $\times$            &   $\times$      &   $\times$   \\ 
\hline
16A-J    &  6   &    0.58   &  0.41  &    0.32    &     0.8 &       0.0 &     0.0  &   2.100e-05   &   $\times$         &    $\times$      &   $\times$            &   $\times$      &   $\times$   \\ 
16A-J    &  6   &    0.96   &  0.04  &    0.60    &     6.1 &       1.7 &    34.7  &   9.014e-05   &     56.3         &      62.1      &     62.1            &   $\times$      &   $\times$   \\ 
16A-J    & 6    &    1.54   &  0.09  &    0.14    &     5.6 &      14.6 &    76.0  &   1.723e-04   &      5.2         &      20.7      &    287.3            &   $\times$      &   $\times$   \\ 
16A-J    &  6   &    1.65   &  0.14  &    0.47    &     4.3 &       0.0 &     0.0  &   2.100e-05   &      5.5         &       8.5      &   $\times$            &   $\times$      &   $\times$   \\ 
16A-J    &  6   &    1.95   &  0.14  &    0.03    &     2.3 &      29.6 &    92.9  &   2.059e-04   &    211.6         &     211.6      &   $\times$            &   $\times$      &   $\times$   \\ 
\hline
17A-J    & 8   &    0.55   &  0.08  &    0.52    &     3.2 &       1.9 &    66.5  &   1.534e-04   &     35.3         &      35.3      &   $\times$            &   $\times$      &   $\times$   \\ 
17A-J    &   8  &    1.12   &  0.07  &    1.03    &    14.1 &       2.9 &    45.4  &   1.114e-04   &      0.2         &      12.8      &     51.8            &    242.2      &   $\times$   \\ 
17A-J    &  8   &    1.63   &  0.30  &    0.05    &     0.7 &       0.0 &     0.0  &   2.100e-05   &   $\times$         &    $\times$      &   $\times$            &   $\times$      &   $\times$   \\ 
17A-J    &  8   &    2.03   &  0.22  &    0.11    &     0.9 &       0.0 &     0.0  &   2.100e-05   &   $\times$         &    $\times$      &   $\times$            &   $\times$      &   $\times$   \\ 
\hline
18A-J    &  10   &    0.32   &  0.14  &    0.05    &     2.2 &      21.6 &    97.1  &   2.143e-04   &     23.8         &      23.8      &   $\times$            &   $\times$      &   $\times$   \\ 
18A-J    &  10   &    0.69   &  0.02  &    1.11    &    16.0 &       3.6 &    53.3  &   1.271e-04   &      3.8         &      11.9      &     32.5            &     44.1      &     95.6   \\ 
18A-J    &  10   &    1.37   &  0.17  &    1.20    &    12.2 &       2.5 &    52.6  &   1.257e-04   &     33.9         &      33.9      &     82.2            &    146.2      &   $\times$   \\ 
18A-J    &  10   &    3.27   &  0.16  &    0.16    &     6.4 &      12.4 &    67.2  &   1.546e-04   &      0.1         &       0.1      &      5.9            &   $\times$      &   $\times$   \\ 
\hline
19A-JS  & 6    &    0.61   &  0.11  &    0.39    &     1.0 &       0.0 &     0.0  &   2.100e-05   &    123.2         &    $\times$      &   $\times$            &   $\times$      &   $\times$   \\ 
19A-JS   & 6    &    0.89   &  0.09  &    0.40    &     5.3 &       2.5 &    39.9  &   1.005e-04   &     63.5         &      82.5      &     91.1            &   $\times$      &   $\times$   \\ 
19A-JS   &  6   &    1.34   &  0.11  &    0.48    &     8.7 &       6.3 &    73.9  &   1.680e-04   &     10.7         &      75.5      &     75.5            &   $\times$      &   $\times$   \\ 
19A-JS   &  6   &    2.65   &  0.08  &    0.05    &     2.8 &      21.6 &    77.1  &   1.745e-04   &      4.0         &       4.0      &   $\times$            &   $\times$      &   $\times$   \\ 
\hline
20A-JS   &  8   &    0.73   &  0.06  &    0.81    &     2.7 &       0.0 &     0.0  &   2.100e-05   &     13.1         &      37.3      &   $\times$            &   $\times$      &   $\times$   \\ 
20A-JS   &  8   &    1.41   &  0.06  &    0.67    &     9.7 &       3.0 &    44.2  &   1.089e-04   &      0.1         &       0.5      &     11.7            &   $\times$      &   $\times$   \\ 
20A-JS   &  8   &    2.18   &  0.22  &    0.04    &     0.3 &       0.0 &     0.0  &   2.100e-05   &   $\times$         &    $\times$      &   $\times$            &   $\times$      &   $\times$   \\ 
20A-JS   &  8   &    3.08   &  0.05  &    0.06    &     1.0 &       0.0 &     0.0  &   2.100e-05   &   $\times$         &    $\times$      &   $\times$            &   $\times$      &   $\times$   \\ 
\hline
21A-JS  &  10   &    0.59   &  0.28  &    0.93    &     2.8 &       0.0 &     0.0  &   2.100e-05   &      4.9         &      24.4      &   $\times$            &   $\times$      &   $\times$   \\ 
21A-JS  & 10    &    1.25   &  0.14  &    0.85    &     9.5 &       2.4 &    44.9  &   1.103e-04   &      8.7         &      27.6      &     52.5            &   $\times$      &   $\times$   \\ 
\hline
\end{tabular}
\end{table}

\begin{table}
\scriptsize
\renewcommand{\arraystretch}{0.8}
\setlength{\tabcolsep}{5pt}
\captcont{Final results of the simulations of model A  (J indicates only Jupiter and JS represents Jupiter-Saturn configuration). From left to right, the columns show  simulation's number, surface density at 1 AU,  semimajor axis,  eccentricity, mass ($M_\oplus$),   amount of water ($O_\oplus$), percentage of asteroidal mass ($> 2.5 AU$), percentage of asteroidal water ($> 2.5 AU$), final D/H ratio, and the time (Myr) of the delivery of $1 O_\oplus$, $2 O_\oplus$, $5 O_\oplus$, $10 O_\oplus$ and $15 O_\oplus$.}
   \centering 
\begin{tabular}{@{}lcccccccccrrrr@{}}
  \hline
  \\
Sim & $\Sigma_1$        & $a_{\rm f}$      & $e_{\rm f}$        & Mass $(M_\oplus)$  & Water $(O_\oplus)$   & \%$M_{\rm ast}$                 &  \%$W_{\rm ast}$ & $D/H$    & $t_{\rm del}$    &  $t_{\rm del}$   &  $t_{\rm del}$   &   $t_{\rm del}$ &  $t_{\rm del}$ \\
     &                &            &              &                      &                        &                       &            &     & $1 O_\oplus$  & $2 O_\oplus$ & $5 O_\oplus$   & $ 10 O_\oplus$ &$ 15 O_\oplus$ \\
\hline\hline
22A-JS   &  6   &    0.60   &  0.04  &    0.37    &     3.0 &       2.7 &    72.1  &   1.645e-04   &     62.0         &      62.0      &   $\times$            &   $\times$      &   $\times$   \\ 
22A-JS   &  6   &    0.79   &  0.11  &    0.36    &     2.2 &       0.0 &     0.0  &   2.100e-05   &      0.6         &      76.7      &   $\times$            &   $\times$      &   $\times$   \\ 
22A-JS   &  6   &    1.31   &  0.02  &    0.55    &     5.4 &       1.8 &    39.6  &   9.982e-05   &     20.1         &      52.7      &    101.5            &   $\times$      &   $\times$   \\ 
22A-JS   &  6   &    2.16   &  0.35  &    0.02    &     0.1 &       0.0 &     0.0  &   2.100e-05   &   $\times$         &    $\times$      &   $\times$            &   $\times$      &   $\times$   \\ 
22A-JS   &  6   &    2.26   &  0.05  &    0.13    &     6.0 &      16.0 &    71.2  &   1.628e-04   &      0.0         &       0.1      &      7.2            &   $\times$      &   $\times$   \\ 
\hline
23A-JS  &  8   &    0.58   &  0.06  &    0.45    &     3.0 &       2.2 &    71.0  &   1.622e-04   &     30.1         &      30.1      &   $\times$            &   $\times$      &   $\times$   \\ 
23A-JS  &  8   &    0.95   &  0.02  &    0.74    &     7.8 &       2.7 &    54.9  &   1.303e-04   &     18.3         &      25.1      &     30.2            &   $\times$      &   $\times$   \\ 
23A-JS  &  8   &    1.41   &  0.04  &    0.58    &     6.9 &       1.7 &    31.1  &   8.295e-05   &     27.1         &      56.0      &     61.1            &   $\times$      &   $\times$   \\ 
\hline
24A-JS   &   10  &    0.55   &  0.12  &    0.47    &     0.6 &       0.0 &     0.0  &   2.100e-05   &   $\times$         &    $\times$      &   $\times$            &   $\times$      &   $\times$   \\ 
24A-JS   &  10   &    0.99   &  0.04  &    1.47    &    12.3 &       0.7 &    17.3  &   5.542e-05   &      6.9         &       9.8      &     16.9            &     43.2      &   $\times$   \\ 
24A-JS   &   10  &    1.87   &  0.20  &    0.12    &     1.4 &       0.0 &     0.0  &   2.100e-05   &    119.4         &    $\times$      &   $\times$            &   $\times$      &   $\times$   \\ 
24A-JS   &  10   &    2.35   &  0.42  &    0.03    &     0.0 &       0.0 &     0.0  &   2.100e-05   &   $\times$         &    $\times$      &   $\times$            &   $\times$      &   $\times$   \\ 
24A-JS   & 10    &    2.51   &  0.25  &    0.11    &     7.7 &      27.9 &    82.8  &   1.857e-04   &      6.9         &       6.9      &    174.4            &   $\times$      &   $\times$   \\ 
\hline
25A-JS   &  6   &    0.60   &  0.11  &    0.45    &     2.7 &       2.2 &    79.3  &   1.788e-04   &      2.1         &       2.1      &   $\times$            &   $\times$      &   $\times$   \\ 
25A-JS   &   6  &    1.09   &  0.03  &    0.60    &     7.0 &       1.7 &    30.3  &   8.138e-05   &     12.9         &      48.1      &    107.5            &   $\times$      &   $\times$   \\ 
\hline
26A-JS   & 8    &    0.60   &  0.18  &    0.86    &     7.0 &       2.3 &    60.9  &   1.423e-04   &     19.5         &      53.0      &    189.5            &   $\times$      &   $\times$   \\ 
26A-JS   &  8   &    1.27   &  0.19  &    0.61    &     8.1 &       3.3 &    53.0  &   1.265e-04   &     16.1         &      16.1      &     40.6            &   $\times$      &   $\times$   \\ 
\hline
27A-JS  &   10  &    0.44   &  0.05  &    0.45    &     0.7 &       0.0 &     0.0  &   2.100e-05   &   $\times$         &    $\times$      &   $\times$            &   $\times$      &   $\times$   \\ 
27A-JS  &  10   &    1.02   &  0.03  &    1.07    &    14.3 &       2.8 &    44.8  &   1.102e-04   &      0.1         &       0.2      &     40.8            &    135.3      &   $\times$   \\ 
27A-JS  &  10   &    1.58   &  0.12  &    0.44    &     6.0 &       4.6 &    70.6  &   1.616e-04   &      7.0         &       7.0      &     12.9            &   $\times$      &   $\times$   \\ 
27A-JS  &  10   &    2.23   &  0.13  &    0.14    &     0.7 &       0.0 &     0.0  &   2.100e-05   &   $\times$         &    $\times$      &   $\times$            &   $\times$      &   $\times$   \\ 
\hline
28A-JS   &  6   &    0.61   &  0.02  &    0.39    &     2.6 &       2.6 &    82.2  &   1.846e-04   &    142.4         &     142.4      &   $\times$            &   $\times$      &   $\times$   \\ 
28A-JS   &   6  &    0.94   &  0.05  &    0.44    &     3.7 &       2.3 &    57.3  &   1.351e-04   &     37.2         &      63.2      &   $\times$            &   $\times$      &   $\times$   \\ 
28A-JS   &  6   &    1.23   &  0.08  &    0.26    &     3.2 &       0.0 &     0.0  &   2.100e-05   &      0.1         &       0.3      &   $\times$            &   $\times$      &   $\times$   \\ 
28A-JS   & 6    &    1.75   &  0.16  &    0.03    &     0.3 &       0.0 &     0.0  &   2.100e-05   &   $\times$         &    $\times$      &   $\times$            &   $\times$      &   $\times$   \\ 
\hline
29A-JS   &  8   &    0.56   &  0.02  &    0.78    &    14.7 &       7.7 &    86.9  &   1.940e-04   &     11.1         &      11.1      &     40.9            &     79.5      &   $\times$   \\ 
29A-JS   &  8   &    1.31   &  0.11  &    0.77    &     6.4 &       1.3 &    33.5  &   8.758e-05   &     13.9         &      13.9      &     45.2            &   $\times$      &   $\times$   \\ 
29A-JS   &  8   &    2.94   &  0.39  &    0.08    &     5.3 &      25.8 &    80.0  &   1.801e-04   &      0.9         &       0.9      &     14.7            &   $\times$      &   $\times$   \\ 
\hline
30A-JS   & 10    &    0.66   &  0.12  &    1.37    &    11.4 &       2.2 &    56.1  &   1.327e-04   &     20.1         &      41.9      &     46.3            &     56.5      &   $\times$   \\ 
30A-JS   &  10   &    1.76   &  0.48  &    0.07    &     1.0 &       0.0 &     0.0  &   2.100e-05   &   $\times$         &    $\times$      &   $\times$            &   $\times$      &   $\times$   \\ 
30A-JS   &  10   &    1.82   &  0.07  &    0.37    &     2.1 &       0.0 &     0.0  &   2.100e-05   &      0.0         &      23.5      &   $\times$            &   $\times$      &   $\times$   \\ 
\hline
31A-JS  &  6   &    0.71   &  0.04  &    0.56    &     6.7 &       3.6 &    64.1  &   1.486e-04   &      9.2         &       9.2      &     52.1            &   $\times$      &   $\times$   \\ 
31A-JS  &  6   &    0.97   &  0.12  &    0.65    &     3.2 &       0.0 &     0.0  &   2.100e-05   &     70.9         &     146.4      &   $\times$            &   $\times$      &   $\times$   \\ 
\hline
32A-JS   &  8   &    0.53   &  0.29  &    0.37    &     0.3 &       0.0 &     0.0  &   2.100e-05   &   $\times$         &    $\times$      &   $\times$            &   $\times$      &   $\times$   \\ 
32A-JS   &  8   &    0.82   &  0.09  &    0.78    &     7.4 &       1.3 &    28.7  &   7.810e-05   &     38.9         &      70.6      &     70.6            &   $\times$      &   $\times$   \\ 
32A-JS   &  8   &    1.55   &  0.04  &    0.48    &     5.3 &       2.1 &    40.4  &   1.013e-04   &     33.2         &      33.2      &     44.1            &   $\times$      &   $\times$   \\ 
32A-JS   &  8   &    3.27   &  0.25  &    0.09    &     9.4 &      42.7 &    90.8  &   2.016e-04   &      0.5         &       0.5      &      0.8            &   $\times$      &   $\times$   \\ 
\hline
33A-JS   &  10   &    0.62   &  0.23  &    1.01    &    12.9 &       5.0 &    82.8  &   1.857e-04   &      2.2         &       2.2      &     28.1            &     73.0      &   $\times$   \\ 
33A-JS   &  10   &    1.38   &  0.05  &    1.20    &    12.9 &       0.8 &    16.5  &   5.384e-05   &      0.9         &       5.5      &     12.5            &     19.9      &   $\times$   \\ 
33A-JS   &  10   &    3.01   &  0.19  &    0.09    &     3.4 &      11.5 &    62.0  &   1.445e-04   &    130.3         &     130.3      &   $\times$            &   $\times$      &   $\times$   \\ 
\hline
34A-JS   &   6  &    0.61   &  0.02  &    0.38    &     2.5 &       2.6 &    83.8  &   1.877e-04   &     75.9         &      75.9      &   $\times$            &   $\times$      &   $\times$   \\ 
34A-JS   &  6   &    1.08   &  0.02  &    0.84    &     7.8 &       1.2 &    27.4  &   7.545e-05   &     26.4         &      31.6      &     98.2            &   $\times$      &   $\times$   \\ 
34A-JS   &  6   &    2.06   &  0.29  &    0.05    &     4.7 &      38.0 &    90.2  &   2.005e-04   &      3.4         &       3.4      &   $\times$            &   $\times$      &   $\times$   \\ 
\hline
\end{tabular}
\end{table}

\begin{table}
\scriptsize
\renewcommand{\arraystretch}{0.8}
\setlength{\tabcolsep}{5pt}
\captcont{Final results of the simulations of model A  (J indicates only Jupiter and JS represents Jupiter-Saturn configuration). From left to right, the columns show  simulation's number, surface density at 1 AU,  semimajor axis,  eccentricity, mass ($M_\oplus$),   amount of water ($O_\oplus$), percentage of asteroidal mass ($> 2.5 AU$), percentage of asteroidal water ($> 2.5 AU$), final D/H ratio, and the time (Myr) of the delivery of $1 O_\oplus$, $2 O_\oplus$, $5 O_\oplus$, $10 O_\oplus$ and $15 O_\oplus$.}
   \centering 
\begin{tabular}{@{}lcccccccccrrrr@{}}
  \hline
  \\
Sim & $\Sigma_1$        & $a_{\rm f}$      & $e_{\rm f}$        & Mass $(M_\oplus)$  & Water $(O_\oplus)$   & \%$M_{\rm ast}$                 &  \%$W_{\rm ast}$ & $D/H$    & $t_{\rm del}$    &  $t_{\rm del}$   &  $t_{\rm del}$   &   $t_{\rm del}$ &  $t_{\rm del}$ \\
     &                &            &              &                      &                        &                       &            &     & $1 O_\oplus$  & $2 O_\oplus$ & $5 O_\oplus$   & $ 10 O_\oplus$ &$ 15 O_\oplus$ \\
\hline\hline
35A-JS   &  8   &    0.18   &  0.24  &    0.11    &     1.9 &       0.0 &     0.0  &   2.100e-05   &      2.8         &    $\times$      &   $\times$            &   $\times$      &   $\times$   \\ 
35A-JS   &   8  &    0.62   &  0.27  &    0.68    &     1.0 &       0.0 &     0.0  &   2.100e-05   &     29.5         &    $\times$      &   $\times$            &   $\times$      &   $\times$   \\ 
35A-JS   &  8   &    1.22   &  0.05  &    1.08    &     8.9 &       0.0 &     0.0  &   2.100e-05   &      3.8         &      35.4      &     52.4            &   $\times$      &   $\times$   \\ 
35A-JS   & 8    &    2.10   &  0.48  &    0.03    &     0.2 &       0.0 &     0.0  &   2.100e-05   &   $\times$         &    $\times$      &   $\times$            &   $\times$      &   $\times$   \\ 
\hline
36A-JS   &   10  &    0.56   &  0.05  &    0.61    &     1.1 &       0.0 &     0.0  &   2.100e-05   &     54.1         &    $\times$      &   $\times$            &   $\times$      &   $\times$   \\ 
36A-JS   &   10  &    1.01   &  0.02  &    1.28    &     9.0 &       0.8 &    23.7  &   6.809e-05   &     14.3         &      14.6      &     17.3            &   $\times$      &   $\times$   \\ 
\hline
\end{tabular}
\end{table}

\clearpage
\section{Final configuration of the simulations  of model B}
 \setcounter{table}{1}
\renewcommand\thetable{\Alph{section}\arabic{table}}

\begin{table}
\scriptsize
\renewcommand{\arraystretch}{0.8}
\setlength{\tabcolsep}{5pt}
\captcont{Final results of the simulations of model B  (J indicates only Jupiter and JS represents Jupiter-Saturn configuration). From left to right, the columns show  simulation's number, surface density at 1 AU,  semimajor axis,  eccentricity, mass ($M_\oplus$),   amount of water ($O_\oplus$), percentage of asteroidal mass ($> 2.5 AU$), percentage of asteroidal water ($> 2.5 AU$), final D/H ratio, and the time (Myr) of the delivery of $1 O_\oplus$, $2 O_\oplus$, $5 O_\oplus$, $10 O_\oplus$ and $15 O_\oplus$.}
   \centering 
\begin{tabular}{@{}lcccccccccrrrr@{}}
  \hline
  \\
Sim & $\Sigma_1$        & $a_{\rm f}$      & $e_{\rm f}$        & Mass $(M_\oplus)$  & Water $(O_\oplus)$   & \%$M_{\rm ast}$                 &  \%$W_{\rm ast}$ & $D/H$    & $t_{\rm del}$    &  $t_{\rm del}$   &  $t_{\rm del}$   &   $t_{\rm del}$ &  $t_{\rm del}$ \\
     &                &            &              &                      &                        &                       &            &     & $1 O_\oplus$  & $2 O_\oplus$ & $5 O_\oplus$   & $ 10 O_\oplus$ &$ 15 O_\oplus$ \\
\hline\hline
1B-J   &  6   &    0.52   &  0.25  &    0.14    &     8.0 &      27.6 &    99.6  &   2.193e-04   &     18.6         &      18.6      &     18.6            &   $\times$      &   $\times$   \\ 
1B-J   &  6   &    0.90   &  0.06  &    0.75    &     2.8 &       0.0 &     0.0  &   2.100e-05   &     33.1         &      33.1      &   $\times$            &   $\times$      &   $\times$   \\ 
1B-J   &  6   &    1.44   &  0.16  &    0.38    &    19.4 &      20.6 &    86.4  &   1.928e-04   &     18.1         &      18.1      &     18.1            &     18.1      &     18.1   \\ 
1B-J   &  6   &    1.46   &  0.45  &    0.03    &     0.5 &       0.0 &     0.0  &   2.100e-05   &   $\times$         &    $\times$      &   $\times$            &   $\times$      &   $\times$   \\ 
1B-J   &  6   &    2.35   &  0.24  &    0.05    &     0.5 &       0.0 &     0.0  &   2.100e-05   &   $\times$         &    $\times$      &   $\times$            &   $\times$      &   $\times$   \\ 
\hline
2B-J  &  8   &    0.59   &  0.18  &    0.56    &     1.6 &       0.0 &     0.0  &   2.100e-05   &     54.2         &    $\times$      &   $\times$            &   $\times$      &   $\times$   \\ 
2B-J  &   8  &    1.01   &  0.14  &    0.45    &     1.6 &       0.0 &     0.0  &   2.100e-05   &     24.1         &    $\times$      &   $\times$            &   $\times$      &   $\times$   \\ 
2B-J  &   8  &    1.39   &  0.05  &    0.88    &    21.6 &       7.8 &    67.9  &   1.562e-04   &      2.8         &      15.7      &     33.5            &     33.5      &     33.5   \\ 
\hline
3B-J   &  10   &    0.47   &  0.51  &    0.19    &     0.0 &       0.0 &     0.0  &   2.100e-05   &   $\times$         &    $\times$      &   $\times$            &   $\times$      &   $\times$   \\ 
3B-J   &  10   &    0.79   &  0.06  &    0.94    &     5.6 &       0.0 &     0.0  &   2.100e-05   &     11.8         &      32.0      &     42.9            &   $\times$      &   $\times$   \\ 
3B-J   & 10    &    1.38   &  0.08  &    1.14    &    29.5 &       7.9 &    65.0  &   1.502e-04   &      0.1         &       1.5      &      6.7            &      6.7      &      6.7   \\ 
3B-J   &  10  &    2.27   &  0.19  &    0.17    &    21.3 &      57.5 &    95.3  &   2.107e-04   &      1.3         &       1.3      &      1.3            &      1.3      &      1.3   \\ 
\hline
4B-J   &   6  &    0.68   &  0.19  &    0.57    &     1.9 &       0.0 &     0.0  &   2.100e-05   &     31.4         &    $\times$      &   $\times$            &   $\times$      &   $\times$   \\ 
4B-J   & 6    &    1.24   &  0.09  &    0.32    &     9.9 &      11.9 &    82.0  &   1.843e-04   &     81.5         &      81.5      &     81.5            &   $\times$      &   $\times$   \\ 
4B-J   &  6   &    1.55   &  0.13  &    0.40    &     3.3 &       0.0 &     0.0  &   2.100e-05   &      2.1         &       7.8      &   $\times$            &   $\times$      &   $\times$   \\ 
4B-J   &  6   &    2.59   &  0.12  &    0.03    &     0.4 &       0.0 &     0.0  &   2.100e-05   &   $\times$         &    $\times$      &   $\times$            &   $\times$      &   $\times$   \\ 
\hline
5B-J   &  8   &    0.72   &  0.16  &    1.22    &    19.4 &       5.6 &    75.6  &   1.714e-04   &     17.6         &      20.3      &     83.6            &     83.6      &     83.6   \\ 
5B-J   &  8   &    1.68   &  0.13  &    0.64    &     5.4 &       0.0 &     0.0  &   2.100e-05   &      1.3         &      28.3      &     66.5            &   $\times$      &   $\times$   \\ 
\hline
6B-J   & 10    &    0.46   &  0.18  &    0.34    &     0.8 &       0.0 &     0.0  &   2.100e-05   &   $\times$         &    $\times$      &   $\times$            &   $\times$      &   $\times$   \\ 
6B-J   &  10   &    0.77   &  0.04  &    0.78    &    20.5 &      10.4 &    84.3  &   1.888e-04   &     27.4         &      27.4      &     27.4            &     27.4      &     27.4   \\ 
6B-J   &  10   &    1.33   &  0.22  &    0.62    &     5.3 &       0.0 &     0.0  &   2.100e-05   &      0.0         &       0.0      &     41.4            &   $\times$      &   $\times$   \\ 
\hline
7B-J   &  6   &    0.62   &  0.20  &    0.60    &     2.8 &       0.0 &     0.0  &   2.100e-05   &    176.5         &     176.5      &   $\times$            &   $\times$      &   $\times$   \\ 
7B-J   &  6   &    1.44   &  0.20  &    0.63    &    20.1 &      12.4 &    82.8  &   1.859e-04   &     38.4         &      69.3      &     69.3            &     69.3      &     69.3   \\ 
7B-J   &  6   &    2.04   &  0.04  &    0.29    &     2.7 &       0.0 &     0.0  &   2.100e-05   &      6.1         &      26.3      &   $\times$            &   $\times$      &   $\times$   \\ 
\hline
8B-J   &  8  &    0.87   &  0.05  &    1.65    &    49.0 &      11.5 &    82.3  &   1.847e-04   &     29.4         &      29.4      &     52.1            &     52.1      &     52.1   \\ 
8B-J   &   8  &    1.59   &  0.10  &    0.46    &     3.6 &       0.0 &     0.0  &   2.100e-05   &      1.6         &       5.3      &   $\times$            &   $\times$      &   $\times$   \\ 
8B-J   &  8   &    2.69   &  0.24  &    0.18    &     2.2 &       0.0 &     0.0  &   2.100e-05   &      0.0         &      49.4      &   $\times$            &   $\times$      &   $\times$   \\ 
\hline
9B-J  &   10  &    0.67   &  0.13  &    1.13    &     3.8 &       0.0 &     0.0  &   2.100e-05   &     18.4         &      21.1      &   $\times$            &   $\times$      &   $\times$   \\ 
9B-J  &  10   &    1.19   &  0.15  &    0.74    &    24.0 &      11.1 &    72.0  &   1.643e-04   &      0.0         &       0.2      &      1.5            &      1.5      &      1.5   \\ 
9B-J  &  10   &    2.14   &  0.19  &    0.65    &     4.2 &       0.0 &     0.0  &   2.100e-05   &     15.1         &      30.9      &   $\times$            &   $\times$      &   $\times$   \\ 
\hline
10B-JS   &  6   &    0.33   &  0.15  &    0.30    &     2.0 &       0.0 &     0.0  &   2.100e-05   &      6.9         &    $\times$      &   $\times$            &   $\times$      &   $\times$   \\ 
10B-JS   &  6   &    0.84   &  0.10  &    0.72    &     2.7 &       0.0 &     0.0  &   2.100e-05   &     58.6         &     150.9      &   $\times$            &   $\times$      &   $\times$   \\ 
10B-JS   &  6   &    1.71   &  0.30  &    0.04    &     0.2 &       0.0 &     0.0  &   2.100e-05   &   $\times$         &    $\times$      &   $\times$            &   $\times$      &   $\times$   \\ 
10B-JS   &   6  &    3.15   &  0.14  &    0.04    &     9.1 &     100.0 &   100.0  &   2.200e-04   &   $\times$         &    $\times$      &   $\times$            &   $\times$      &   $\times$   \\ 
\hline
11B-JS  &  8   &    0.58   &  0.10  &    0.54    &     0.8 &       0.0 &     0.0  &   2.100e-05   &   $\times$         &    $\times$      &   $\times$            &   $\times$      &   $\times$   \\ 
11B-JS  &   8  &    0.93   &  0.18  &    0.34    &     1.3 &       0.0 &     0.0  &   2.100e-05   &     41.5         &    $\times$      &   $\times$            &   $\times$      &   $\times$   \\ 
11B-JS  &  8   &    1.43   &  0.11  &    0.78    &     5.9 &       0.0 &     0.0  &   2.100e-05   &      2.4         &      30.1      &     85.3            &   $\times$      &   $\times$   \\ 
11B-JS  &  8   &    2.18   &  0.11  &    0.12    &    25.2 &     100.0 &   100.0  &   2.200e-04   &      0.9         &       0.9      &      0.9            &      0.9      &      0.9   \\ 
\hline
12B-JS   &  10   &    0.94   &  0.05  &    2.26    &    38.1 &       4.3 &    54.3  &   1.290e-04   &      7.5         &      17.7      &     31.2            &     31.2      &     31.2   \\ 
12B-JS   &  10   &    1.55   &  0.13  &    0.27    &     1.0 &       0.0 &     0.0  &   2.100e-05   &      5.6         &    $\times$      &   $\times$            &   $\times$      &   $\times$   \\ 
\hline
13B-JS   &   6  &    0.58   &  0.17  &    0.39    &     0.7 &       0.0 &     0.0  &   2.100e-05   &   $\times$         &    $\times$      &   $\times$            &   $\times$      &   $\times$   \\ 
13B-JS   &  6   &    0.98   &  0.10  &    0.63    &     3.6 &       0.0 &     0.0  &   2.100e-05   &     30.5         &      71.6      &   $\times$            &   $\times$      &   $\times$   \\ 
13B-JS   &  6   &    2.73   &  0.02  &    0.04    &     9.1 &     100.0 &   100.0  &   2.200e-04   &   $\times$         &    $\times$      &   $\times$            &   $\times$      &   $\times$   \\ 
\hline
\end{tabular}
\end{table}

\begin{table}
\scriptsize
\renewcommand{\arraystretch}{0.8}
\setlength{\tabcolsep}{5pt}
\captcont{Final results of the simulations of model B  (J indicates only Jupiter and JS represents Jupiter-Saturn configuration). From left to right, the columns show  simulation's number, surface density at 1 AU,  semimajor axis,  eccentricity, mass ($M_\oplus$),   amount of water ($O_\oplus$), percentage of asteroidal mass ($> 2.5 AU$), percentage of asteroidal water ($> 2.5 AU$), final D/H ratio, and the time (Myr) of the delivery of $1 O_\oplus$, $2 O_\oplus$, $5 O_\oplus$, $10 O_\oplus$ and $15 O_\oplus$.}
   \centering 
\begin{tabular}{@{}lcccccccccrrrr@{}}
  \hline
  \\
Sim & $\Sigma_1$        & $a_{\rm f}$      & $e_{\rm f}$        & Mass $(M_\oplus)$  & Water $(O_\oplus)$   & \%$M_{\rm ast}$                 &  \%$W_{\rm ast}$ & $D/H$    & $t_{\rm del}$    &  $t_{\rm del}$   &  $t_{\rm del}$   &   $t_{\rm del}$ &  $t_{\rm del}$ \\
     &                &            &              &                      &                        &                       &            &     & $1 O_\oplus$  & $2 O_\oplus$ & $5 O_\oplus$   & $ 10 O_\oplus$ &$ 15 O_\oplus$ \\
\hline\hline
14B-JS &  8   &    0.83   &  0.09  &    1.56    &     8.0 &       0.0 &     0.0  &   2.100e-05   &      8.1         &      26.0      &     47.4            &   $\times$      &   $\times$   \\ 
14B-JS &   8  &    2.74   &  0.15  &    0.06    &    12.7 &     100.0 &   100.0  &   2.200e-04   &   $\times$         &    $\times$      &   $\times$            &   $\times$      &   $\times$   \\ 
\hline
15B-JS  &  10   &    0.64   &  0.10  &    1.13    &     6.4 &       0.0 &     0.0  &   2.100e-05   &      2.2         &      19.9      &     55.3            &   $\times$      &   $\times$   \\ 
15B-JS  &  10   &    1.20   &  0.27  &    1.19    &    25.2 &       7.0 &    69.9  &   1.600e-04   &      0.0         &       0.6      &     12.3            &     12.3      &     12.3   \\ 
15B-JS  &  10   &    3.04   &  0.17  &    0.09    &    19.7 &     100.0 &   100.0  &   2.200e-04   &   $\times$         &    $\times$      &   $\times$            &   $\times$      &   $\times$   \\ 
\hline
16B-JS  & 6    &    0.68   &  0.22  &    0.64    &    18.7 &      12.1 &    87.6  &   1.952e-04   &     20.7         &      75.4      &     75.4            &     75.4      &     75.4   \\ 
16B-JS  & 6    &    1.26   &  0.18  &    0.49    &     2.3 &       0.0 &     0.0  &   2.100e-05   &     41.9         &      61.2      &   $\times$            &   $\times$      &   $\times$   \\ 
16B-JS  &  6  &    3.28   &  0.17  &    0.09    &    18.9 &     100.0 &   100.0  &   2.200e-04   &      0.0         &       0.0      &      0.0            &      0.0      &      0.0   \\ 
\hline
17B-JS   &  8   &    0.62   &  0.05  &    0.69    &     3.5 &       0.0 &     0.0  &   2.100e-05   &     71.8         &      71.8      &   $\times$            &   $\times$      &   $\times$   \\ 
17B-JS   &   8  &    0.97   &  0.05  &    0.60    &    15.4 &       9.7 &    80.7  &   1.815e-04   &     26.5         &      26.5      &     26.5            &     26.5      &    146.6   \\ 
17B-JS   &  8   &    1.59   &  0.23  &    0.25    &     1.5 &       0.0 &     0.0  &   2.100e-05   &     17.1         &    $\times$      &   $\times$            &   $\times$      &   $\times$   \\ 
\hline
18B-JS   &  10   &    0.76   &  0.10  &    1.76    &    42.4 &       9.8 &    86.3  &   1.928e-04   &     12.8         &      33.2      &     55.8            &     55.8      &     55.8   \\ 
18B-JS   &  10   &    2.51   &  0.08  &    0.42    &    60.9 &      66.1 &    96.5  &   2.131e-04   &      0.7         &       0.7      &      0.9            &      0.9      &      0.9   \\ 
\hline
\end{tabular}
\end{table}

\label{lastpage}


\begin{thebibliography}{99}
\bibitem[Abe et al.(2000)]{2000orem.book..413A} Abe, Y., Ohtani, E., Okuchi, T., Righter, K., \& Drake, M., 2000, Origin of 
the Earth and Moon, 413
\bibitem[Agnor et al.(1999)]{1999Icar..142..219A} Agnor, C.~B., Canup, R.~M., \& Levison, H.~F., 1999, Icarus, 142, 219 
\bibitem[Albar{\`e}de(2009)]{2009Natur.461.1227A} Albar{\`e}de, F.\ 2009, 
Nature, 461, 1227

\bibitem[Asaduzzaman 
\& Muralidharan(2012)]{2012M&PSA..75.5081A} Asaduzzaman, A.~M., \& Muralidharan, K.\ 2012, Meteoritics and Planetary Science Supplement, 75, 5081 

\bibitem[Bethell 
\& Bergin(2009)]{2009Sci...326.1675B} Bethell, T., \& Bergin, E.\ 2009, Science, 326, 1675 

\bibitem[Biver et al.(2006)]{2006A&A...449.1255B} Biver, N., et al.\ 2006, A\&A, 449, 1255 


\bibitem[biver et al.(1989)]{biver} Biver, N., Bockelée-Morvan, D., Boissier, J., et al. 2005, Asteroids, 
Comets, Meteors 2005, IAU Symp. No 229, abstract book, 43

\bibitem[Bockelee-Morvan et al.(1998)]{1998Icar..133..147B}Bockelee-Morvan, D., Gautier D., Lis D. C., Young K., Keene J., 
Phillips T., Owen T., Crovisier J., Goldsmith P. F., Bergin E. A., Despois D., Wootten A., 1998, Icarus, 133, 147 
\bibitem[Boss(1996)]{1996ApJ...469..906B} Boss, A.~P.\ 1996, ApJ, 469, 906 

\bibitem[\protect\citeauthoryear{Campins et al.}{2004}]{b5} Campins H., Swindle T. D., Kring D. A., 2004, Origin, 
Evolution and Biodiversity of Microbial Life in the Universe, 569
\bibitem[Canup 
\& Pierazzo(2006)]{2006LPI....37.2146C} Canup, R.~M., \& Pierazzo, E., 2006, 37th Annual Lunar and Planetary Science Conference, 37, 2146 


\bibitem[\protect\citeauthoryear{Chambers}{1999}]{b6} Chambers J. E., 1999, MNRAS, 304, 793
\bibitem[\protect\citeauthoryear{Chambers}{2001}]{b7} Chambers J. E., 2001, Icarus, 152, 205

\bibitem[Chambers 
	\& Wetherill(1998)]{1998Icar..136..304C} Chambers, J.~E., \& Wetherill, G.~W.\ 1998, Icarus, 136, 304 





\bibitem[Clark(1998)]{1998epsn.conf.....C} Clark, S.\ 1998, Extrasolar 
planets : the search for new worlds / Stuart Clark.~ New York : Wiley, 
1998.~(Wiley-Praxis series in astronomy and astrophysics) QB820 .C53 1998,  

\bibitem[\protect\citeauthoryear{Dauphas et al.}{2000}]{b9} Dauphas N., Robert F., Marty B., 2000, Icarus, 148, 508


\bibitem[\protect\citeauthoryear{Deloule et al.}{1998}]{b10} Deloule E., Robert F., Doukhan J. C., 1998, GCA, 62, 3367

\bibitem[Delsemme(1998)]{1998P&SS...47..125D} Delsemme, A.~H.\ 1998, Planet. Spance Sci., 47, 125

\bibitem[Drake(2005)]{2005M&PS...40..519D} Drake, M.~J.\ 2005, Meteoritics and Planetary Science, 40, 519 

\bibitem[Drake \& Campins(2006)]{2006IAUS..229..381D} Drake, M.~J., \& Campins, H., 2006, Asteroids, Comets, Meteors, 229, 381 


\bibitem[\protect\citeauthoryear{Drake \& Righter}{2002}]{b12}Drake M. J., Righter K., 2002, Nature, 416, 39
\bibitem[\protect\citeauthoryear{Innanen}{1991}]{b13} Drouart A., Dubrulle B., Gautier D., Robert F., 1999,
Icarus, 140, 129
\bibitem[\protect\citeauthoryear{Eberhardt et al.}{1995}]{b31} Eberhardt P., Reber M., Krankowsky D., Hodges R. R.,
1995, AAP, 302, 301
\bibitem[Eisner(2007)]{2007Natur.447..562E} Eisner, J.~A., 2007, Nature, 447,562
\bibitem[Encrenaz(2006)]{2006swu..book.....E} Encrenaz, T.\ 2006, Searching 
for Water in the Universe, by T.~Encrenaz.~ISBN 978-0-387-34174-3.~ Berlin: 
Springer, 2006.
\bibitem[Fegley(2000)]{2000SSRv...92..177F} Fegley, B., Jr.\ 2000, Space Science Reviews, 
92, 177 
\bibitem[Genda \& Abe(2005)]{2005Natur.433..842G} Genda, H., \& Abe, Y., 2005, Nature, 433, 842 
\bibitem[Genda \& Ikoma(2008)]{2008Icar..194...42G} Genda, H., \& Ikoma, M., 2008, Icarus, 194, 42
\bibitem[\protect\citeauthoryear{Gomes et al.}{2005}]{b15} Gomes R., Levison H. F., Tsiganis K., Morbidelli A., 2005, Nature, 435, 466
\bibitem[\protect\citeauthoryear{Murray et al.}{2007}]{b16} Haghighipour N., Raymond S. N., 2007, ApJ, 666, 436

\bibitem[Haghighipour 
\& Scott(2012)]{2012ApJ...749..113H} Haghighipour, N., \& Scott, E.~R.~D.\ 2012, \apj, 749, 113 


\bibitem[Haghighipour et al.(2005)]{2012CMDA.....}  Haghighipour  N., Izidoro A., Winter O. C. 2013, submitted to CMDA

\bibitem[Hansen(2009)]{2009ApJ...703.1131H} Hansen, B.~M.~S.\ 2009, \apj, 
703, 1131 




\bibitem[Hartogh et 
al.(2009)]{2009P&SS...57.1596H} Hartogh, P., et al.\ 2009, Planetary and Space Science
, 57, 1596 

\bibitem[Hartogh et al.(2011)]{2011Natur.478..218H} Hartogh, P., Lis, 
D.~C., Bockel{\'e}e-Morvan, D., et al.\ 2011, \nat, 478, 218 

	\bibitem[Hartogh et al.(2011)]{2011Natur.478..218H} Hartogh, P., Lis, 
	D.~C., Bockel{\'e}e-Morvan, D., et al.\ 2011, Nature, 478, 218 

\bibitem[Hayashi (1981)]{Hayashi81}
Hayashi, C., 1981, Prog. Theor. Phys. Suppl., 70, 35


\bibitem[Hirschmann et al.(2005)]{2005E&PSL.236..167H} Hirschmann, M.~M., Aubaud, C.,
	 \& Withers, A.~C.\ 2005, Earth and Planetary Science Letters, 236, 167 


Haghighipour


\bibitem[Hutsem{\'e}kers et al.(2009)]{2009Icar..204..346H}Hutsem{\'e}kers, D., Manfroid, J., Jehin, E., 
\& Arpigny, C., 2009, Icarus, 204, 346

\bibitem[Hutsem{\'e}kers et al.(2008)]{2008A&A...490L..31H} Hutsem{\'e}kers, D.,
	 Manfroid, J., Jehin, E., Zucconi, J.-M., \& Arpigny, C.\ 2008, A\&A, 490, L31 

\bibitem[Ikoma 
\& Genda(2006)]{2006ApJ...648..696I} Ikoma, M., \& Genda, H.\ 2006, \apj, 648, 696 

\bibitem[Ipatov 
\& Mather(2007)]{2007IAUS..236...55I} Ipatov, S.~I., \& Mather, J.~C.\ 2007, IAU Symposium, 236, 55 

\bibitem[Izidoro et al(2013)]{000} Izidoro, A., Haghighipour, N., Winter O. C., Tsuchida M., under revision ApJ, 2013


\bibitem[Jacobsen(2005)]{2005AREPS..33..531J} Jacobsen, S.~B.\ 2005, Annual 
Review of Earth and Planetary Sciences, 33, 531 



\bibitem[Jessberger et al.(1989)]{1989oeps.book..167J} Jessberger, E.~K., Kissel, J., 
\& Rahe, J., 1989, Origin and Evolution of Planetary and Satellite Atmospheres, 167 


\bibitem[Kasting et al.(1993)]{1993Icar..101..108K} Kasting, J.~F., 
Whitmire, D.~P., \& Reynolds, R.~T., 1993, Icarus, 101, 108 

 

\bibitem[\protect\citeauthoryear{Lecuyer \& Gillet}{1998}]{b22}Lecuyer C., Gillet P., Robert F., 1998, Chem. Geol., 145,249
\bibitem[\protect\citeauthoryear{Lellouch et al. }{2001}]{b23}Lellouch E., Bézard B., Fouchet T., Feuchtgruber H., 
Encrenaz T., de Graauw T., 2001, AAP, 370, 610
 
\bibitem[Levison 
\& Agnor(2003)]{2003AJ....125.2692L} Levison, H.~F., \& Agnor, C.\ 2003, AJ, 125, 2692 



\bibitem[\protect\citeauthoryear{Lunine}{2003}]{b25} Lunine J. I., Chambers J., Morbidelli A., Leshin L. A., 2003, Icarus, 165, 1
\bibitem[Marty(2012)]{2012E&PSL.313...56M} Marty, B.\ 2012, Earth and Planetary Science Letters, 313, 56

\bibitem[Marty (2006)]{Marty (2006)}Marty, B., Yokochi, R., 2006, Rev. Mineral. Geochem. 62, 421

\bibitem[Matsui 
\& Abe(1986)]{1986Natur.322..526M} Matsui, T., \& Abe, Y., 1986, Nature, 322, 526 



\bibitem[\protect\citeauthoryear{Meyer et al.}{1998}]{b27} Meier R., Owen T. C., Jewitt D. C., Matthews H. E., Senay M., 
Biver N., Bockelee-Morvan D., Crovisier J., Gautier D., 1998, Science, 279, 1707


\bibitem[Morbidelli et al.(2012)]{2012AREPS..40..251M} Morbidelli, A., 
Lunine, J.~I., O'Brien, D.~P., Raymond, S.~N., 
\& Walsh, K.~J.\ 2012, Annual Review of Earth and Planetary Sciences, 40, 251 

\bibitem[Morbidelli et 
al.(2000)]{2000M&PS...35.1309M} Morbidelli, A., Chambers, J., Lunine, J.~I., Petit, J.~M., Robert, F., Valsecchi, G.~B., 
\& Cyr, K.~E., 2000, Meteoritics and Planetary Science, 35, 1309 




\bibitem[Muralidharan et al.(2008)]{2008Icar..198..400M} Muralidharan, K., Deymier, P., Stimpfl, M., de Leeuw, N.~H., 
\& Drake, M.~J., 2008, Icarus, 198, 400


\bibitem[Nuth(2008)]{2008EM&P..102..435N} Nuth, J.~A.\ 2008, Earth Moon and Planets, 102, 435 

\bibitem[\protect\citeauthoryear{O'Brien et al.}{2006}]{b30} O'Brien D. P., Morbidelli A., Levison H. F., 2006, Icarus, 184, 39



\bibitem[Owen \& Bar-Nun(1995)]{1995Icar..116..215O} Owen, T., \& Bar-Nun, A.\ 1995, Icarus, 116, 215
 
\bibitem[Owen 
\& Bar-Nun(2000)]{2000orem.book..459O} Owen, T.~C., \& Bar-Nun, A.\ 2000, Origin of the Earth and Moon, 459 



\bibitem[\protect\citeauthoryear{Raymond et al.}{2004}]{b35} Raymond S. N., Quinn T., Lunine J. I., 2004, Icarus, 168, 1
\bibitem[\protect\citeauthoryear{Raymond et al.}{2005}]{b36}Raymond S. N., Quinn T., Lunine J. I., 2005a, Icarus, 177, 256
\bibitem[Raymond et al.(2005)]{2005ApJ...632..670R} Raymond, S.~N., Quinn, T., \& Lunine, J.~I., 2005b, Apj, 632, 670 
\bibitem[Raymond et al.(2007)]{2007ApJ...669..606R} Raymond, S.~N., Scalo, J., \& Meadows, V.~S., 2007, Apj, 669, 606 
\bibitem[\protect\citeauthoryear{Raymond et al.}{2007}]{b37} Raymond S. N., Quinn T., Lunine J. I., 2007, Astrobiology, 7, 66
\bibitem[Raymond et al.(2006)]{2006Icar..183..265R} Raymond, S.~N., Quinn, T., \& Lunine, J.~I.\ 2006, Icarus, 183, 265 
\bibitem[Raymond et al.(2009)]{2009Icar..203..644R} Raymond, S.~N., 
O'Brien, D.~P., Morbidelli, A., \& Kaib, N.~A., 2009, Icarus, 203, 644 


\bibitem[Rubie et 
al.(2011)]{2011E&PSL.301...31R} Rubie, D.~C., Frost, D.~J., Mann, U., et al.\ 2011, Earth and Planetary Science Letters, 301, 31 



\bibitem[Sasselov \& Lecar(2000)]{2000ApJ...528..995S} Sasselov, D.~D., \& Lecar, M.\ 2000, ApJ, 528, 995 
\bibitem[Schmidt et al.(2005)]{2005AGUFM.P11A0102S} Schmidt, B.~E., Brown, 
R.~H., \& Lauretta, D.~S., 2005, AGU Fall Meeting Abstracts, A102 

\bibitem[Smyth et al.(2006)]{2006GeoRL..3315301S} Smyth, J.~R., Frost, 
D.~J., Nestola, F., Holl, C.~M., \& Bromiley, G., 2006, Geophysical Research Letters, 33, 15301
 
\bibitem[Stimpfl et al.(2004)]{2004M&PSA..39.5218S} Stimpfl, M., Lauretta, D.~S., \& Drake, M.~J.\ 2004, Meteoritics 
and Planetary Science Supplement, 39, 5218 

\bibitem[Stimpfl et al.(2006)]{2006JCrGr.294...83S} Stimpfl, M., Walker, A.~M., Drake, M.~J., de Leeuw, N.~H., \& Deymier, P.,  2006, Journal of Crystal Growth, 294, 83

\bibitem[Swindle \& Kring(2001)]{2001eag..conf.3785S} Swindle, T.~D., \& Kring, D.~A.\ 2001, Eleventh Annual V.~M.~Goldschmidt 
Conference, 3785

\bibitem[Touboul et al.(2007)]{2007Natur.450.1206T} Touboul, M., Kleine, T., Bourdon, B., Palme, H., \& Wieler, R.\ 2007, Nature, 450, 1206 

\bibitem[Villanueva et al.(2009)]{2009ApJ...690L...5V} Villanueva, G.~L., 
Mumma, M.~J., Bonev, B.~P., Di Santi, M.~A., Gibb, E.~L., B{\"o}hnhardt, 
H., \& Lippi, M.\ 2009, ApJL, 690, L5 

\bibitem[Walsh et al.(2011)]{2011Natur.475..206W} Walsh, K.~J., Morbidelli, 
A., Raymond, S.~N., O'Brien, D.~P., \& Mandell, A.~M.\ 2011, \nat, 475, 206 

\bibitem[Weaver et al.(2008)]{2008LPICo1405.8216W} Weaver, H.~A., A'Hearn, 
	M.~F., Arpigny, C., et al.\ 2008, LPI Contributions, 1405, 8216 

\bibitem[Weidenschilling (1977)]{Weidenschilling77}
Weidenschilling, S. J., 1977a, Astrophys. Space Sci., 51, 153

\bibitem[Weirich et al.(2004)]{2004DPS....36.3301W} Weirich, J.~R., Brown, R.~H., \& Lauretta, D.~S.\ 2004, 
Bulletin of the American Astronomical Society, 36, 1143 

\bibitem[Wetherill(1991)]{1991LPI....22.1495W} Wetherill, G.~W.\ 1991, 
Lunar and Planetary Institute Science Conference Abstracts, 22, 1495 

\bibitem[Williams 
\& Hemley(2001)]{2001AREPS..29..365W} Williams, Q., \& Hemley, R.~J.,  2001, Annual Review of Earth and Planetary Sciences, 29, 365 

\bibitem[Wood et al.(2008)]{2008GeCoA..72.1415W} Wood, B., Wade, J., 
\& Kilburn, M.\ 2008, \gca, 72, 1415 


\end{thebibliography}
\end{document}